\documentclass[aps,preprint,showpacs,showkeys]{revtex4}
\usepackage{graphicx}
\usepackage{amsmath}
\usepackage{amsfonts}
\usepackage{amsthm}
\usepackage{mathrsfs}
\usepackage{amssymb}
\usepackage{graphicx}
\usepackage{epsfig}
\usepackage{hyperref}
\usepackage{amscd}
\usepackage{epstopdf}
\begin{document}
\newcommand{\be}{\begin{equation}}
\newcommand{\ee}{\end{equation}}
\newcommand{\bea}{\begin{eqnarray}}
\newcommand{\eea}{\end{eqnarray}}

\title{APPROXIMATE PERFECT FLUID SOLUTIONS WITH QUADRUPOLE MOMENT}

\author{ Medeu Abishev$ {^{* \pounds}}$, Nurzada Beissen ${^{* \dag}}$, Farida Belissarova $ {^{* \pounds}}$, Kuantay Boshkayev $^{* \ddag \mathsection }$, Aizhan Mansurova $^*$, 
Aray Muratkhan $^{* \pounds}$, Hernando Quevedo $^{* \P \rVert}$ , Saken Toktarbay $^{* \pounds}$}

\affiliation{ $^*$ Institute \ for \ Experimental \ and \ Theoretical \ Physics \\ Al-Farabi \ Kazakh National University \\ 71 al-Farabi Ave., Almaty, 050040, Kazakhstan, $^ \dag$ M.Kh. Dulaty \ Taraz \ Regional \ University
\\ st. Tole Bi 40, Taraz, 080000, Kazakhstan, $^\ddag$ National Nanotechnology Laboratory of Open Type \\ Al-Farabi \ Kazakh National University \\ 71 al-Farabi Ave., Almaty, 050040, Kazakhstan, $^\mathsection$ Satbayev \ Kazakh \ National Technical \ University \\ Satbayev St. 22a, Almaty, 050000, Kazakhstan,
$^ \P$ Instituto \ de \ Ciencias \ Nucleares \\ Universidad \ Nacional \ Aut\'onoma \ de \ M\'exico \\ AP 70543, M\'exico, D.F. 04510, Mexico, $^\rVert$ Dipartimento \ di \ Fisica \ and \ ICRA \\ Universit\`a di Roma ``La Sapienza" \\ I-00185 Roma, Italy,
$^\pounds$ Institute of Nuclear Physics \\ 1 Ibragimova St.
Almaty,050032, Kazakhstan, 
$^{*}$ saken.yan@yandex.com }

\date{}

\begin{abstract}
We investigate the interior Einstein's equations in the case of a static, axially symmetric, perfect fluid source. We present a particular line element that is specially suitable for the investigation of this type of interior gravitational fields. Assuming that the deviation from spherically symmetry is small, we linearize the corresponding line element and field equations and find several classes of vacuum and perfect fluid solutions. We find some particular approximate solutions  by imposing appropriate matching conditions.
\end{abstract}


\maketitle



\section{\label{sec:intro}Introduction}

Based on experimental evidence, general relativity is considered today as one of the best candidates  to describe the gravitational field of compact astrophysical objects. As a theory for the gravitational field, it should be able to describe the field of all possible physical configurations, in which gravity is involved. All the information about the gravitational field should be contained in the metric tensor which must be a solution of Einstein's equations. Consider the case of a compact object like a star or a planet. From the point of view of the multipole structure of the source, to describe the field of a compact object, we need an interior and an exterior solution, both containing at least  three independent physical  parameters, namely, mass, angular momentum and quadrupole moment.
Consider first the case of a  source with only mass and angular momentum. The corresponding exterior solution is represented by the 
Kerr spacetime \cite{kerr63} for which no physically reasonable interior solution is known.  This is a major problem in classical general relativity \cite{solutions}.  Many methods have been suggested to find a  suitable interior Kerr solution, including exotic matter models and specially adapted equations of state,  but none of them has lead to a definite answer;  in this regard, see \cite{hph17} for a recent promising proposal. In view of this situation, we consider that alternative approaches should be considered. In particular, we believe that additional physical parameters can be taken into account that are relevant for the description of the gravitational field. The simplest of such additional parameters is the quadrupole moment which is responsible for the deformation of any realistic mass distribution. Indeed, if we add a quadrupole moment to a spherically symmetric object, we end up with an axisymmetric mass distribution, which implies new degrees of freedom at the level of the corresponding field equations. This is the main idea of the alternative approach we propose to attack the problem of finding interior solutions to describe the interior gravitational structure of compact objects. As a first step to develop such an approach, we will focus in this work on the case of a source with only mass and quadrupole, neglecting the contribution of the angular momentum.    
In a recent work \cite{quev11}, it was proposed to use the Zipoy-Voorhees transformation \cite{zip66,voor70} to generate the quadrupolar metric ($q-$metric), which can be interpreted as the simplest generalization of the Schwarzschild metric, describing the gravitational field of a distribution of mass whose non-spherically symmetric shape is represented by an independent quadrupole parameter. {In the literature, this metric is known as the Zipoy-Voorhees metric, delta-metric, gamma-metric and $q$-metric \cite{mala04,Mashhoon2018}. Here, we will use the name $q$-metric to highlight the importance of the quadrupole parameter $q$}. Indeed, several studies show the physical importance of the quadrupole parameter. 
Circular and radial geodesics of the exterior gamma-metric ($\gamma=1+q$) have been compared with the spherically symmetric case to establish the sensitivity of the trajectories to the gamma parameter \cite{Herrera2000IJMPD}. Moreover, it was shown that the properties of the accretion disks in the field of the gamma-metric can be drastically different from those of disks around black holes
\cite{Chowdhury2012,Quevedo2016,Boshkayev2016}.
These studies show that the $q-$metric can be used to describe the exterior gravitational field of deformed distributions of mass in which the quadrupole moment is the main parameter that describes the deformation. 

The question arises whether it is possible to find an interior metric that can be matched to the exterior one in such a way that the entire spacetime is described as a whole. 
 To this end, it is usually assumed that the interior mass distribution can be described by means of a perfect fluid with two physical 
 parameters, namely, energy density and pressure. The energy-momentum tensor of the perfect fluid is then used in the Einstein equations as the source of
 the gravitational field.  
 It turns out that the system of  corresponding differential equations cannot be solved because the number of equations is less than the number of unknown functions. This problem is usually solved by imposing equations of state that relate the pressure and density of the fluid. Moreover, one should impose
 energy conditions, matching conditions with the exterior metric, and conditions on the behavior of the metric functions near the center 
 of the source and on the boundary with the exterior field.    
 
 {Hernandez \cite{Hernandez1967}  has shown how  to modify {\it ad hoc} an interior  spherically symmetric solution  to  obtain an approximate interior solution  for the corresponding family of exterior Weyl metrics, provided the exterior metric contains the  Schwarzschild metric as a particular case. The Hernandez approach has been generalized by Stewart et al. \cite{Stewart1982} to obtain an exact interior  solution  to the gamma-metric. They found two different interior  solutions which  match the exterior  gamma-metric. In general, however, this {\it ad hoc} method does not lead to interior solutions corresponding to simple fluids.
 The matching between interior and exterior solutions, in general, requires the fulfillment of several mathematical conditions on the matching surface \cite{Senovilla93,Herrera05}.
 
 The above discussion shows that the search for physically relevant interior solutions is not an easy task. The difficulties increase once we consider the non-uniqueness of the solutions. Indeed, whereas the Birkhoff theorem guarantees that the Schwarzschild metric is the only spherically symmetric vacuum solution of Einstein's equations, there exist many spherically symmetric interior solutions that can be matched with the Schwarzschild metric. In the case of axial symmetry,  the situation is even more complicated. In a recent work \cite{2018RSOS....570826F,Boshkayevpxa}, several exterior solutions with quadrupole were compared. It was found that they are all represented by diverse analytical expressions and  are characterized by different sets of multipole moments. In this sense, they are all physically different from each other. One can, therefore, expect that there will be many interior metrics that can be matched with each one of the exterior solutions. One example of this situation is given by a recently proposed interior solution for the $q-$metric \cite{Hernandez_Pastora_2016} and the solutions that we will analyze in this work. In fact, to obtain the solution found in \cite{Hernandez_Pastora_2016} the authors have chosen a particular line element which does not contain the line element used in this work. For the sake of completeness, we show this result in \ref{app:hphm}.

  This work is organized as follows. In Sec. \ref{sec:exq}, we consider the $q-$metric as describing the exterior gravitational field of a deformed source with mass and quadrupole  moment.
         In Sec. \ref{sec:int}, we present the exact field equations for a perfect fluid source and the corresponding conservation laws, which take a particular simple form for the chosen line element.     Then, in Sec. \ref{sec:appsol},  
  we construct the approximate interior and exterior line elements, in which the quadrupole moment is considered up to the first order only. We analyze in detail the field equations for the exterior field and derive a 5-parametric approximate solution, which we interpret as the most general vacuum solution with arbitrary mass and small quadrupole moment. We also consider the approximate field equations and matching conditions for a perfect fluid. We find the spherically symmetric metric with vanishing quadrupole, which is used in the following sections as background solution.  
In Sec. \ref{sec:sol}, we find several particular interior solutions. First, we fix some metric functions as constants and show that there exist perfect fluid solutions with well-behaved pressure and density, but with discontinuities in the derivatives of the metric functions, which is usually considered as a nonphysical
matching behavior. Then, we postulate a particular radial function for the density and find numerical solutions, which satisfy the matching conditions for the metric functions and the energy conditions for the fluid. Finally, we consider a barotropic equation of state (EoS) and find well-behaved solutions. 
The physical significance of the solutions is investigated in Sec. \ref{sec:phys}. In particular, we show that some solutions are characterized by a physically meaningful behavior of pressure and density, which is essentially comparable with the behavior obtained from more sophisticated EoSs for white dwarfs and neutron stars. This shows that essentially the numerical solutions obtained in this work can be applied to describe the gravitational field of realistic compact objects. 
   Finally, in Sec. \ref{sec:con}, we  discuss of our results.

\section{\label{sec:exq}Exterior q-metric}

 Zipoy \cite{zip66} and Voorhees \cite{voor70} investigated static, axisymmetric vacuum solutions of Einstein's equations and found a simple transformation, which allows one to generate new solutions from a known solution. 
If we start from the Schwarzschild solution and apply a Zipoy-Voorhees transformation, the new line element can be written as
\cite{quev11}
\label{sec:exqmet}

\begin{eqnarray} \nonumber
ds^2 =&&  \left(1-\frac{2m}{r}\right)^{1+q} dt^2- \left(1-\frac{2m}{r}\right)^{-q} \nonumber \\
 & &\times \bigg[ \left(1+\frac{m^2\sin^2\theta}{r^2-2mr}\right)^{-q(2+q)}\left(\frac{dr^2}{1-\frac{2m}{r}}+ r^2d\theta^2\right) \nonumber \\
 & & + r^2 \sin^2\theta d\varphi^2\bigg].
\label{exqmet}
\end{eqnarray} 
This metric is axially symmetric and, consequently, it is physically different from the seed Schwarzschild metric. 
A detailed analysis shows that $m$ and   $q$ are constant parameters that determine the total mass and the quadrupole moment of the gravitational source  \cite{quev11}. A stationary generalization of the $q-$metric, satisfying the main physical conditions of exterior spacetimes, has
been obtained in \cite{saken14q,2018RSOS....580640F}.
The metric (\ref{exqmet}) has been interpreted as the simplest generalization of the Schwarzschild metric with a quadrupole. 

As mentioned above, whereas the  seed metric is the spherically symmetric Schwarzschild solution, which describes the gravitational field of a black hole, the generated $q-$metric is axially symmetric, and describes the exterior field of a naked singularity \cite{quev11}.
In fact, this can be shown explicitly by calculating the invariant Geroch multipoles 
\cite{GerochI,GerochII}. The 
lowest mass multipole moments $M_n$, $n=0,1,\ldots $ are given by
\begin{equation}
M_0= (1+q)m\ , \quad M_2 = -\frac{m^3}{3}q(1+q)(2+q)\ ,
\end{equation}
whereas higher moments are proportional to $mq$ and can be 
completely rewritten in terms of $M_0$ and $M_2$. Accordingly, the arbitrary parameters $m$ and $q$ determine the mass and quadrupole 
which are the only independent multipole moments of the solution. In the limiting case $q=0$, only the monopole $M_0=m$ 
survives, as in the Schwarzschild spacetime. In the limit $m=0$, with $q\neq 0$, all moments vanish identically, implying that 
no mass distribution is present and the spacetime must be flat. The same is true in the limiting case $q\rightarrow -1$ which corresponds
to the Minkowski metric. Moreover, notice that all odd multipole moments are zero because the solution possesses an additional 
reflection symmetry with respect to the equatorial plane $\theta=\pi/2$.  

The deformation is described by the quadrupole moment $M_2$ which is positive for a prolate source 
and negative for an oblate source. This implies that the parameter $q$ can be either positive or negative. 
Since the total mass $M_0$ of the source must be positive, we must assume that $q>-1$ for positive values of $m$, and $q<-1$ for negative values of $m$. We conclude that the above metric can be used to describe the exterior gravitational field of a static positive mass $M_0$ with a positive or negative quadrupole moment $M_2$. 

A study of the curvature of the $q-$metric shows that the outermost singularity is located at $r=2m$, a hypersurface which in all known  compact objects is situated inside the surface of the body. This implies that in order to describe the entire gravitational field, it is necessary to cover this type of singularity with an interior solution.


\section{\label{sec:int}Interior metric}

As mentioned in the Introduction, in this work, we will concentrate on the case of static perfect fluid spacetimes. There are many forms to write down the corresponding line element and, in principle, all of them could lead to different particular solutions \cite{solutions}. Therefore, the choice of the line element is important for obtaining particular families of solutions. Certain forms of the line element turn out to be convenient for investigating a particular problem. Our experience with numerical 
perfect fluid solutions \cite{quev12} indicates that for the case under consideration the line element 
\begin{equation}
ds^2 = fdt^2 - \frac{e^{2\gamma}}{f}\left(\frac{dr^2}{h} + d\theta^2\right) -\frac{\mu^2}{f}d\varphi^2\  ,
\label{in_lel}
\end{equation}
is particularly convenient.  Here $f=f(r,\theta)$, $\gamma=\gamma(r,\theta)$, $\mu=\mu(r,\theta)$, and $h=h(r)$. 
A redefinition of the coordinate $r$ leads to an equivalent line element which has been used to investigate anisotropic static fluids \cite{hdio13}. 

The Einstein equations for a perfect fluid with 4-velocity $U_\alpha$, density $\rho$, and pressure $p$ (we use geometric units with $G=c=1$)
\begin{equation}
R_{\alpha\beta} - \frac{1}{2} R g_{\alpha\beta} = 8 \pi \left[(\rho+p)U_\alpha U_\beta - p g_{\alpha\beta}\right] 
\label{eins}
\end{equation}
for the line element (\ref{in_lel}) can be represented as two second-order differential equations for $\mu$ and $f$
\begin{equation}
\mu_{,rr} = -\frac{1}{2h} \left( 2 \mu_{, \theta\theta} + h_{,r} \mu_{,r} - 32 \pi p\frac{\mu e^{2\gamma}}{f} \right) \ ,\label{eqmu}
\end{equation}

\begin{eqnarray}
f_{,rr} =&& \frac{f_{,r}^2}{f} -\left(\frac{h_{,r}}{2h} + \frac{\mu_{,r}}{\mu}\right)f_{,r} + \frac{f_{,\theta}^2}{hf} -
\frac{\mu_{,\theta} f_{,\theta}}{\mu h} -\frac{f_{,\theta\theta}}{h} \nonumber \\ && + 8\pi \frac{(3p+\rho)e^{2\gamma}}{h}\ ,\label{eqf}
\end{eqnarray}
where a subscript represents partial derivative. Moreover, 
the function $\gamma$ is determined by a set of two partial differential equations

\begin{eqnarray} \nonumber
\gamma_{,r}=&& \frac{1}{h\mu_{,r}^2 + \mu_{,\theta}^2} \bigg\{\frac{\mu}{f^2}\bigg[\frac{\mu_{,r}}{4}\left(h f_{,r}^2-f_{,\theta}^2\right) 
+\frac{1}{2}\mu_{,\theta}f_{,\theta}f_{,r} \\&& + 8\pi\mu_{,r} p f e^{2\gamma}\bigg] + \mu_{,\theta}\mu_{,r \theta}-\mu_{,r}\mu_{,\theta\theta} \bigg\},
\label{gamr}
\end{eqnarray}

\begin{eqnarray} \nonumber
\gamma_{,\theta} =&&\frac{1}{h\mu_{,r}^2+\mu_{,\theta}^2}\bigg\{ \frac{ \mu}{f^2} \bigg[ \frac{\mu_{,\theta}}{4}\left(f_{,\theta}^2-h f_{r}^2\right) 
+\frac{1}{2} h \mu_{,r} f_{,\theta}f_{,r}  \nonumber \\ && - 8\pi\mu_{,\theta} p f e^{2\gamma} \bigg] 
+ h\mu_{,r}\mu_{,r\theta} +\mu_{,\theta}\mu_{,\theta\theta}\bigg\}\ ,
\label{gamt}
\end{eqnarray}
which can be integrated by quadratures once $f$, $\mu$,  $p$, and $h$ are known. The integrability condition of these partial
differential equations turns out to be satisfied identically by virtue of the remaining field equations. Notice that there is no 
equation for the function $h(r)$. This means that it can be absorbed in the definition of the radial coordinate $r$. Nevertheless, 
one can also fix it arbitrarily; it turns out that this freedom is helpful when solving the equations and investigating the physical significance of the solutions. 

The advantage of using the line element (\ref{in_lel}) is that the field equations are split into two sets. 
The main set consists of the equations (\ref{eqmu}) and (\ref{eqf}) for $\mu$ and $f$ which must be solved simultaneously. The second set consists of the first-order equations for $\gamma$ which plays a secondary role in the sense that they can be integrated once the remaining functions are known. Notice also that the pressure $p$ and the density $\rho$ must be given {\it a priori} in order to solve the main set of differential equations for $\mu$ and $f$.  As follows from Eq.(\ref{eqf}), the equation of state $3p+\rho=0$ reduces the complexity of this equation; nevertheless, this condition leads to negative pressures which, from a physical point of view, are not expected to be present inside astrophysical compact objects. 

Finally, we mention that from the conservation law $T^{\alpha\beta}_{ \ ;\beta} =0$, we obtain two first-order differential equations for the pressure
\begin{equation}
p_{,r} = - \frac{1}{2} (p+\rho) \frac{f_{,r}}{f}\ , \quad p_{,\theta} =  - \frac{1}{2} (p+\rho) \frac{f_{,\theta}}{f}\ ,
\label{eqp}
\end{equation}
that can be integrated for any given functions $f(r,\theta)$ and $\rho(r,\theta)$, which satisfy Einstein's equations. 

 It is very difficult to find physically reasonable solutions for the above field equations, because the underlying differential equations are highly nonlinear with very strong couplings between the metric functions. In \cite{SH_generate},  some of us
presented a new method for generating perfect fluid solutions of the Einstein
equations, starting from a given seed solution. The method is based upon the introduction of a new
parameter at the level of the metric functions of the seed solution in such a way that the generated
new solution is characterized by physical properties which are different from those of the seed
solutions. 

In this work, we will analyze approximate solutions which satisfy the conditions for being applicable in the case of 
astrophysical compact objects. We will see that it is then possible to perform a numerical integration by imposing
appropriate initial conditions. In particular, if we demand that the metric functions and the pressure are finite 
at the axis, it is possible to find a class of numerical solutions which can be matched with the exterior $q-$metric 
with a pressure that vanishes at the matching surface.
\section{\label{sec:appsol}Linearized quadrupolar metrics}

Our general goal is to investigate how the quadrupole moment influences the structure of spacetimes that can be used to
describe the gravitational field of compact deformed gravitational sources. In particular, we aim
to find perfect fluid solutions that can be matched with the exterior $q-$metric (\ref{exqmet}). 
   
To find the corresponding interior line element, we proceed as follows. Consider the case of a slightly deformed mass. 
This means that the parameter $q$ for the exterior $q-$metric can be considered as small and we can
linearize the line element as

\begin{eqnarray} \nonumber
ds^2 = &&\left(1-\frac{2m}{r}\right) \left[ 1+ q\ln\left(1-\frac{2m}{r}\right)\right] dt^2 \nonumber \\ & & 
- r^2 \bigg[ 1- q\ln\left(1-\frac{2m}{r}\right) \bigg] \sin^2\theta d\varphi^2 \nonumber \\ & &
- \bigg[ 1 + q  \ln \left(1-\frac{2m}{r}\right) - 2q \ln \bigg(1-\frac{2m}{r} \nonumber \\ && + \frac{m^2}{r^2} \sin^2\theta \bigg)\bigg] \left(\frac{dr^2}{1-\frac{2m}{r}} + r^2 d\theta^2\right)
\label{qmatch}
\end{eqnarray}
We will assume that the exterior gravitational field of the compact object is described to the first order in $q$ by the line element 
(\ref{qmatch}), which represents a particular approximate solution to Einstein's equations in vacuum.

To construct the approximate interior line element, we start from the exact line element (\ref{in_lel}) and use the approximate solution 
(\ref{qmatch}) as a guide.  
Following this procedure, an appropriate interior line element can be expressed as
\begin{eqnarray}
ds^2 =&&  e^{2 \nu} (1+qa) dt^2 - (1+qc+qb)\frac{dr^2}{1-\frac{2\tilde m}{r}} \nonumber \\ &&-
(1+qa+qb)r^2d\theta^2
-(1-qa)r^2\sin^2\theta d\varphi^2\ ,
\label{apin1}
\end{eqnarray} 
where the functions $\nu=\nu(r)$, $a=a(r)$, $c=c(r)$, $\tilde m=\tilde m(r)$, and $b=b(r,\theta)$. Notice that we have introduced an additional auxiliary function $c(r)$ which plays a role similar to that of the auxiliary function $h(r)$ of the interior line 
element (\ref{in_lel}). Notice that the approximate line element (\ref{apin1}) contains also the approximate exterior $q-$metric (\ref{qmatch})
as a particular case. This implies that vacuum fields can also  be investigated by using this approximate line element.
\subsection{\label{sec:gvs}General vacuum solution}

To test the consistency of the linearized approach, we will derive explicitly the approximate vacuum $q-$metric (\ref{qmatch}).
To this end, we compute the vacuum field equations from the line element (\ref{apin1}) and obtain
\begin{equation}
\tilde m_{,r} = 0 \quad {\rm i.e.} \quad \tilde m = m = const.\ ,
\end{equation}

\begin{equation}
\nu_{,r}=\frac{ {m}}{r \left(r-2 {m} \right)}\ ,
\label{nuv}
\end{equation}

\begin{equation}
(r-m)(a_{,r}-c_{,r}) + (a-c) = 0\ ,
\label{acv}
\end{equation}

\begin{eqnarray} \nonumber
2 &&r \left(r-2m \right)a_{,rr} + \left(3r-m \right) a_{,r} + \left(r-3m \right) c_{,r} \\ &&- 2 \left(a-c \right)=0,
\end{eqnarray} 

\begin{eqnarray} \nonumber
r &&\left(r-2m \right) b_{,rr}+ b_{,\theta \theta} + \left(r-m \right) b_{,r} -2 \left(r-2m \right) c_{,r} \\ &&+ 2 \left(a-c \right)=0 ,
\end{eqnarray} 

\begin{eqnarray} \nonumber 
&& \left( r^2 -2 mr + m^2 \sin^2 \theta  \right) b_{,\theta} + 2 r \left(r-2m \right) \nonumber \\ && \times \left( m a_{,r}-  a+c \right) \sin \theta \cos \theta =0 ,
\end{eqnarray}

\begin{eqnarray} \nonumber 
&&\left( r^2 -2 m r + m^2 \sin^2 \theta  \right)   b_{,r}+ 2 \left(r-2 m \right) \nonumber \\ && \times \left(r- m \sin^2 \theta \right) a_{,r}+2 \left(r-m \right) \left(a-c \right) \sin^2 \theta =0 ,
\end{eqnarray}
where for simplicity we have replaced the solution of the first equation $\tilde m = m = const.$ in the remaining equations.

Then, Eqs.(\ref{nuv}) and (\ref{acv}) can be integrated and yield
\begin{equation}
\nu = \frac{1}{2} \ln\left(1-\frac{2m}{r}\right) + \alpha_1 \ ,\quad  a-c = \frac{\alpha_2 m^2}{(r-m)^2} \ ,
\end{equation}
where $\alpha_1$ and $\alpha_2$ are dimensionless integration constants. 
The remaining system of partial differential equations can be integrated in general and yields

\begin{equation}
a= - \frac{\alpha_2 m}{r-m} + \frac{1}{2}\left(\alpha_3 - {\alpha_2}\right) \ln\left(1-\frac{2m}{r}\right) + \alpha_4\ ,
\label{vsola}
\end{equation}

\begin{equation}
c= - \frac{\alpha_2 m r}{(r-m)^2} + \frac{1}{2}\left(\alpha_3 - {\alpha_2}\right) \ln\left(1-\frac{2m}{r}\right) + \alpha_4\ ,
\label{vsolc}
\end{equation}

\begin{eqnarray} \nonumber
b=&& \frac{2\alpha_2 m}{r-m} -  \left(\alpha_3 - {\alpha_2}\right) 
\bigg[ \ln 2 \\ && + \ln\left(1-\frac{2m}{r} + \frac{m^2\sin^2\theta}{r^2} \right)\bigg] + \alpha_5 \ ,
\label{vsolb}
\end{eqnarray}
where $\alpha_3$, $\alpha_4$ and $\alpha_5$ are dimensionless integration constants. 

Thus, we see that the general approximate exterior solution with quadrupole moment is represented by a 5-parameter family of solutions. The particular case 

\begin{eqnarray} \nonumber
&&\alpha_1= 0\ , \quad \alpha_2 = 0\ , \quad \alpha_3 = 2\ , \quad \alpha_ 4 = 0\ , \\ && \quad \alpha_5= 2 \ln 2 \ ,
\end{eqnarray} 
corresponds to the linearized $q-$metric as represented in Eq.(\ref{qmatch}). Another interesting particular case corresponds to the choice
\begin{equation} 
\alpha_1 =0 \ ,\quad \alpha_3 = {\alpha_2}\ ,\quad \alpha_4= 0\ ,\quad \alpha_5 = 0\ ,
\end{equation}
which leads to the following line element

\begin{eqnarray} \nonumber
ds^2 &= & \left(1-\frac{2m}{r}\right)\left(1-\frac{q\alpha_2 m}{r-m}\right) dt^2 \nonumber \\ & & - \left(1+  \frac{q \alpha_2 m}{r-m}\right) r^2 \sin^2\theta d\varphi^2  \nonumber \\ & & -\left[1+  \frac{q\alpha_2 m (r-2m)}{(r-m)^2} \right] \frac{dr^2}{1-\frac{2m}{r}} \nonumber \\ & & - \left(1 + \frac{q\alpha_2 m }{r-m}\right) r^2d\theta^2
\label{appsol1}
\ .
\end{eqnarray}
This is an asymptotically flat approximate solution with parameters $m$, $q$ and $\alpha_2$. The singularity structure can be found by analyzing the Kretschmann invariant $K= R_{\alpha\beta\gamma\delta} R^{\alpha\beta\gamma\delta}$  which in this case reduces to 
\begin{equation}
K= \frac{48 m^2}{r^6}\left(1+ q \alpha_2 \frac{r-4m}{r-m} + \mathcal{O}(q^2)\right)\ ,
\end{equation}
where the term proportional to $q^2$ has been neglected due to the approximate character of the solution. We see that there is a central singularity at $r=0$ and a second one at $r=m$. We conclude that the solution (\ref{appsol1})  describes the exterior field of two naked 
singularities of mass $m$ and quadrupole $q$.   
The parameter $\alpha_2$ can be absorbed by redefining the constant $q$ and so it has no special physical meaning.
In the general solution (\ref{vsola})-(\ref{vsolb}), the additive constants $\alpha_4$ and $\alpha_5$ can be chosen such that at infinity the solution describes the Minkowski spacetime in spherical coordinates. This means that non asymptotically flat solutions are also contained 
in the $5-$parameter family (\ref{vsola})-(\ref{vsolb}). This is the most general vacuum solution which is linear in the quadrupole moment. To our knowledge, this general solution is new.

\subsection{\label{sec:NutLimit}Newtonian limit}

To further investigate the physical meaning of the solution (\ref{appsol1}), let us consider  the coordinate transformations \cite{Mashhoon2018}
\begin{equation}
r = R \bigg[1- q \frac{m}{R} \left( 1+ \frac{m}{R} \left( \beta_{1} + \sin^2 \vartheta \right) + \frac{m^2}{R ^2} \left( \beta_{2} - \sin^2 \vartheta \right)+ .. \right) \sin^2 \vartheta \bigg],
\label{trans_r}
\end{equation}
and
\begin{equation}
\theta = \vartheta -q \frac{m^2}{R^2} \left(1+ 2 \frac{m}{R} + . . . \right) \sin \vartheta \cos \vartheta
\label{trans_theta},
\end{equation}
where the $\beta_1$ and $\beta_2$ are  constants and we have neglected terms older than $m^3 / R^3 $. Inserting the above coordinates into the metric (\ref{appsol1}), we obtain the approximate line element
\begin{equation}
ds^2= \left( 1+ {2 \Phi} \right)dt^2 - \frac{dR^2 }{1+ {2 \Phi}}  - U \left( R, \vartheta \right) R^2 \left( d\vartheta^2 + \sin^2 \vartheta d \varphi^2 \right), 
\label{Newline}
\end{equation}
with  
\begin{equation}
    \Phi= -\frac{GM}{R}+ \frac{GQ}{R^3} P_{2}\left( \cos \vartheta \right),
\end{equation}
\begin{equation}
    U \left( R, \vartheta \right)= 1- 2 \frac{GM}{R^3} P_2(\cos \vartheta),
\end{equation}
where $P_2(\cos \vartheta)$ is the Legendre polynomial of degree 2, and we have chosen the free constants as $\alpha_{2}=2$, $\beta{_1}=1/3 $ and $\beta{_2}=5/3 $.

We recognize the metric (\ref{Newline}) as the Newtonian limit of general relativity, where $\Phi$ represents the Newtonian potential. Moreover, the constants  
\begin{equation}
    M= \left( 1+q \right)m, \ \ \ Q=\frac{2}{3} q m^3,
\end{equation}
can be interpreted as the Newtonian mass and quadrupole moment of the corresponding mass distribution. 

We conclude that the metric (\ref{appsol1}) represents the exterior gravitational field of a slightly deformed mass. We will use this exterior approximate metric to match the interior solutions we will investigate in the following section. 

Notice that in order to obtain the Newtonian limit (\ref{Newline}), it is necessary to apply the coordinate transformation (\ref{trans_r}), which relates $r$ with $R$ and $\vartheta$. This shows that $r$ cannot interpreted as a radial coordinate and surfaces with $r=const$ do not correspond to spheres.

\section{\label{sec:pfs}Perfect fluid solutions}

We now apply the approximate line element (\ref{apin1}) to the study of perfect fluid solutions. First, we note that in this case
the conservation law (\ref{eqp}) reduces to
\begin{equation}
p_{,r} = -(\rho+p) \nu_{,r}\ ,\qquad p_{,\theta} = 0\ .
\label{eqp1}
\end{equation}
Calculating the second derivative $p_{,r\theta}=0$, the above conservation laws lead to
\begin{equation}
\rho_{,\theta}=0\ ,
\label{rhotheta}
\end{equation}
implying that the perfect fluid variables can depend on the coordinate $r$ only. As mentioned in the previous section, this does not  imply that the source is spherically symmetric. In fact, due to the presence of the quadrupole parameter $q$ in the line element (\ref{apin1}), the coordinate $r$ is no longer a radial coordinate  and the equation $r=$constant represents, in general, a non-spherically symmetric deformed surface 
\cite{zn71}.

The corresponding linearized Einstein equations can be represented as 
\begin{equation}
 \stackrel{(0)}{G_ {\mu }^{\nu}}+ q \stackrel{(q)}{G_ {\mu }^{\nu}}=8 \pi \left(\stackrel{(0)}{T_ {\mu }^{\nu}}+q \stackrel{(q)}{T_ {\mu }^{\nu}}\right),
\label{l_ein}
\end{equation}
where the $(0)-$terms correspond to the limiting case of spherical symmetry. As for the energy-momentum tensor, we assume that density and pressure can also be linearized as 
\begin{equation}
 p(r)=p_0(r)+q  p_1(r), \ \ \rho(r)=\rho_0(r)+q\rho_1(r),
\label{eqstate} 
 \end{equation}
in accordance with the conservation law conditions (\ref{eqp1}) and (\ref{rhotheta}). Here, $p_0(r)$ and $\rho_0(r)$ are the pressure and
density of the background spherically symmetric solution, respectively. If we now compute the linearized field equations (\ref{l_ein}) for the line element  (\ref{apin1}), we arrive at a set of nine differential equations for the functions $\nu,\ \tilde m,\ a,\ b,\  c,$ $\rho_1$ and $p_1$.
After lengthy computations, it is then possible to isolate an equation that relates $p_1(r)$ and $b(r,\theta)$ from which it follows
that 
\begin{equation}
b_{,\theta}=0 \ .
\end{equation}
This means that for the particular approximate line element (\ref{apin1}), the field equations for a perfect fluid do not allow the metric functions to explicitly depend on the angular coordinate $\theta$. To search for concrete solutions which can be matched with an exterior metric with quadrupole, it is necessary to modify the exterior metric accordingly. Therefore, we will now consider the alternative
approximate metric (\ref{appsol1}), with  $\alpha_{2}=2$, which can be expressed as 
\begin{eqnarray} \nonumber
ds^2 &= & \left(1-\frac{2m}{r}\right)\left(1-\frac{2q m}{r-m}\right) dt^2 \nonumber \\ & & - \left(1+  \frac{2q  m}{r-m}\right) r^2 \sin^2\theta d\varphi^2  \nonumber \\ & & -\left[1+  \frac{2q m (r-2m)}{(r-m)^2} \right] \frac{dr^2}{1-\frac{2m}{r}} \nonumber \\ & & - \left(1 + \frac{2q m }{r-m}\right) r^2d\theta^2\ .
\label{appsol_1}
\end{eqnarray}
As shown in the previous section, this approximate solution leads to the Newtonian limit (\ref{Newline}) and can be used to describe the exterior field of a slightly deformed mass.

We will see that this approximate exterior solution can be used together with the interior line element (\ref{apin1}) to search for 
approximate solutions with a perfect fluid source. Taking into account that the conservation laws and the approximate 
field equations for a perfect fluid imply that the physical quantities  $p$ and $\rho$ and the metric function $b$ depend only on the spatial coordinate $r$, the remaining field equations can be represented explicitly as given in  
\ref{app_linfield}. 

\subsection{\label{sec:back}The background solution}

For the zeroth component of the linearized field equations, we will consider a spherically symmetric spacetime. 
If we set $q=0$ in 
the line element (\ref{apin1}), only the metric functions $\nu$ and $\tilde m$ remain for which  we obtain the field equations
\begin{equation}
\tilde{m}_{,r} =4 \pi r^2 \rho_0,
\end{equation}
\begin{equation}
\nu_{,r} = \frac{\tilde{m}+4 \pi r^3 p_0 }{r \left( r-2 \tilde{m}\right)}.
\end{equation}  
If we assume that the density is constant, $\rho_0=$ const., we obtain a particular solution that can be represented as 
\begin{eqnarray} \nonumber
&& e^{\nu}= e^{\nu_0}=\frac{3}{2}  f_0 (R) - \frac{1}{2}  f _0(r) , \ \tilde m = \frac{4\pi}{3} \rho_0 r^3, \\ && p_0 = \rho_0 \frac { f_0 (r) - f _0 (R)}{ 3 f _0 (R) - f_0 (r)}, 
\label{int_sch} \ 
\end{eqnarray}
with 
\begin{equation}
f_0 (r) = \sqrt{ 1 - \frac{2mr^2}{R^3}},
\label{int_sch_f}
\end{equation}
where the integration constants have been chosen such that at the surface radius $r=R$ the exterior Schwarzschild metric is obtained. The resulting line element
\begin{eqnarray} \nonumber
ds^2 =&& \frac{1}{4}[3f_0(R) -f_0(r)]^2 dt^2 - \frac{dr^2}{1-\frac{8\pi}{3}\rho_0 r^2} \\ &&- r^2(d\theta^2+\sin^2\theta d\varphi^2)
\end{eqnarray}
represents the simplest spherically symmetric perfect fluid solution and is known as the interior Schwarzschild metric. In this work, we will use it as the zeroth approximation of the interior quadrupolar solutions to be obtained below.

\subsection{\label{sec:match}Matching conditions}

The importance of writing the approximate line elements as given above is that the matching between the interior and the exterior 
metrics can be performed in a relatively easy manner. Indeed, let us consider the boundary conditions at the matching surface $r=r_{\Sigma}$  by comparing the above
interior metric (\ref{apin1}) with the $q-$metric (\ref{appsol_1}) to first order in $q$. Then, we obtain the matching conditions
\begin{eqnarray} \nonumber
&& a(r_{\Sigma})=-\frac{2m}{r_{\Sigma}-m},\ \ c(r_{\Sigma}) =-\frac{2m}{\left( r_{\Sigma}-m \right)^2} \ \  b(r_{\Sigma})= \frac{4m}{r_{\Sigma}-m} + \alpha_{5} , \\  &&  \nu(r_{\Sigma}) = \frac{1}{2} \ln \left( 1- \frac{2m}{r_{\Sigma}} \right) \ \  \tilde m (r_{\Sigma}) =m.
\label{bouncona} 
\end{eqnarray}

In addition, we can impose the physically meaningful condition that the total pressure vanishes at the matching surface, i.e.,
\begin{equation}
p(r_{\Sigma}) = 0 \ .
\label{bounconp}
\end{equation}
From the point of view of a numerical integration, the above matching conditions can be used as boundary values for the 
integration of the  corresponding differential equations. 

Notice that we reach the desired matching by fixing only the spatial coordinate as $r=r_{\Sigma} $; however, as mentioned in the previous section, this does not mean that the matching 
surface is a sphere. Indeed, the shape of the matching surface is determined by the conditions $t=const$ and $r=r_{\Sigma} $ which, according to Eq.(\ref{Newline}),  determine a surface with explicit $\theta-$dependence. The coordinate $r$ is therefore not a radial coordinate.  This has been previously observed in the case of a different metric with quadrupole moment \cite{zn71}.

\section{\label{sec:sol}Particular interior solutions}

Our goal is to find interior solutions to the linearized system of differential equations which take into account the contribution of the quadrupole parameter $q$ only up to the first order.  The explicit form of the corresponding  field equations is given in 
\ref{app_linfield}. One can see that they can be split into two sets that can be treated separately. The first set relates only the functions
 $\tilde m(r)$, $\nu(r)$ and $\rho_0(r)$, which must
satisfy the equations
\begin{equation}
\tilde m_{,r} = 4 \pi \rho_0 r^2\ ,
\label{Schm} 
\end{equation}
\begin{equation}
\nu_{,r}=\frac{4 \pi p_{0} r^3 + \tilde{m}}{r \left(r-2 \tilde{m} \right)}\ ,
\label{Schv}
\end{equation}
and
\begin{equation}
p_{0,r}=-\frac{\left( 4 p_{0} \pi r^3+\tilde{m} \right) \left(p_{0}+ \rho_{0} \right)}{r\left(r-2 \tilde{m}\right)} \ .
\label{Schp0}
\end{equation}
This set of equations can be integrated immediately once the value of the density $\rho_0$ is known. 
In particular, for a constant 
$\rho_0$, we obtain the interior Schwarzschild metric (\ref{int_sch}), which is the zeroth order solution we will use in the following sections to integrate the field equations. 
 
In addition, for the remaining functions $a(r)$, $b(r)$ and $c(r)$, we obtain a set of two second-order and three first-order partial differential equations which are presented explicitly in \ref{app_linfield}. In the next sections, we will analyze this set of equations and derive several particular solutions.

\subsection{\label{sec:fcons}Solutions determined by constants}

To find an interior counterpart for the  approximate exterior $q-$metric, 
we consider first the simplest case in which $a$, $b$, $c$ and $\rho_{1}$ are constants. 
The explicit form of the remaining field equations 
(see \ref{app_linfield}) suggests the 
relationship
\begin{equation}
b=-2 a + C_{ab}\ , \quad C_{ab} = const.\  ,
\end{equation}
which reduces considerably the complexity of the equations. Indeed, the only non-trivial equations in this case are
\begin{equation}
\rho_{1}=- \left(b+c\right)\rho_0\ ,\ \
p_{1}= -\left(b+c\right) p_{0},
\end{equation}
so that the total pressure and density are 
\begin{eqnarray} \nonumber
&& p\left(r \right)= p_{0}(r) \left[1 -q \left(b+c \right) \right], \\ && \rho= \rho_{0} \left[1 -q \left(b+c \right) \right]\ ,
\end{eqnarray}
where $p_0(r)$ is given in Eq.(\ref{int_sch}). 
The values of the constants $a$, $b$ and $c$ can be determined from the matching 
conditions with the exterior metric on the surface $r=r_{\Sigma}$. We obtain   
\begin{eqnarray} \nonumber
&& \nu \left(r_{\Sigma} \right)=-1.020, \tilde m \left(r_{\Sigma} \right) = 0.435 \ ,\quad a\left(r_\Sigma \right) =-1.540,\ \  c \left(r_{\Sigma} \right) = -2.725, \ \\ && b\left( r_\Sigma \right) = 3.080, \ C_{ab}=0.01 \label{B_bounconb}. 
\end{eqnarray}

This is a simple approximate interior solution in which the presence of the quadrupole parameter essentially leads to a modification of the pressure of the body. For instance, for the particular choice 
\begin{eqnarray} 
\label{con_parmet}
\rho_0=\frac{3 m}{4 \pi r_{\Sigma}^3}, \ R=1,\ m=0.435, \ \ q=\frac{1}{100}\ ,
\end{eqnarray}
we obtain the pressure and the metric functions depicted in Fig.\ref{fig1}. 
\begin{figure}
\includegraphics[width=6.5cm,height=6cm,scale=0.1]{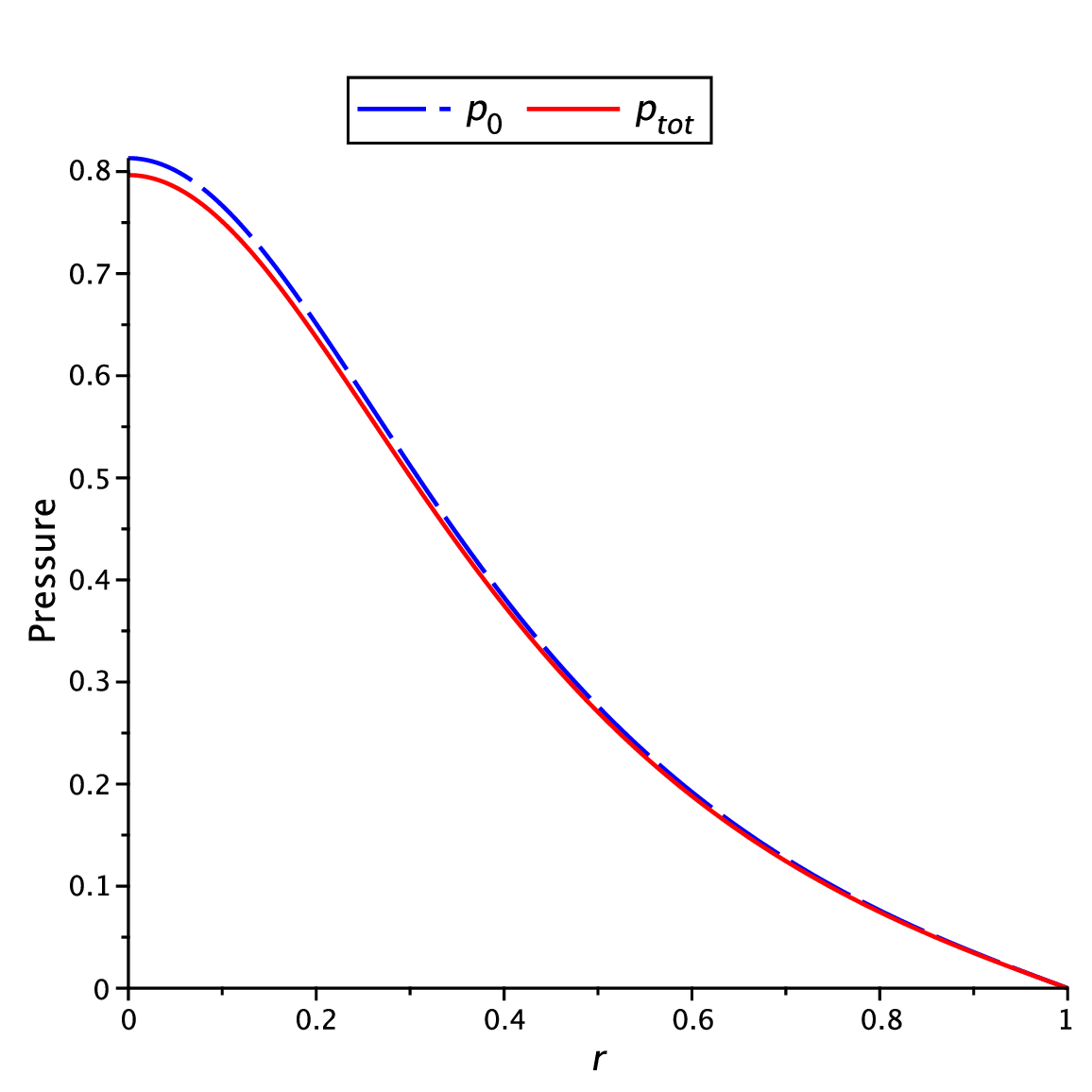}
\includegraphics[width=6.5cm,height=6cm,scale=0.1]{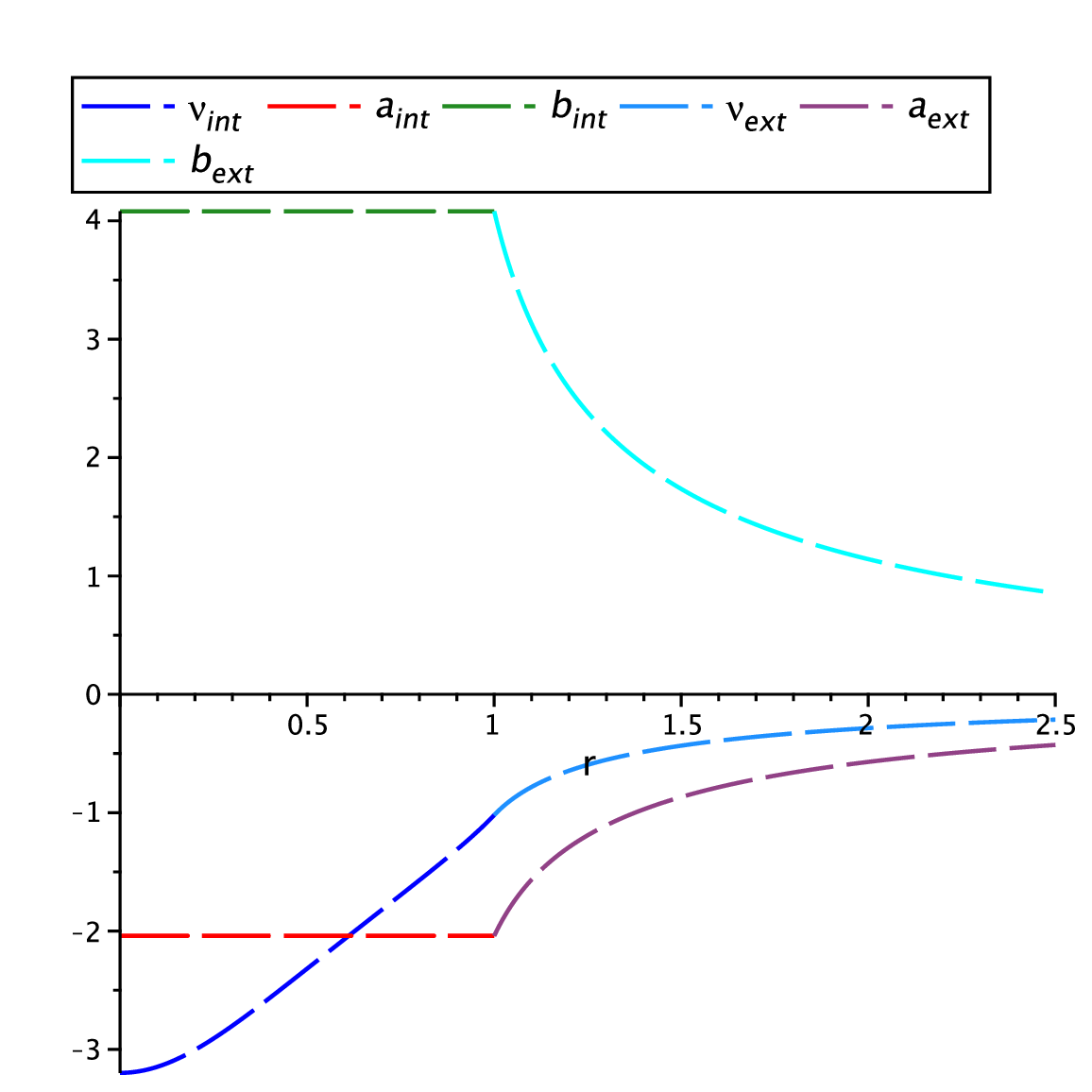}
\caption{Behavior of the pressure and the metric functions in terms of the spatial coordinate $r$ in units of $m$. 
\label{fig1} 
 }
\end{figure}
We conclude that this interior solution is singularity free and can be matched continuously across the matching surface $r=r_{\Sigma}$  with the approximate exterior metric (\ref{appsol_1}).

Notice, however, the discontinuity in the derivatives of the metric functions. This means that the simple Ansatz of a constant interior metric is indeed compatible with the field equations, but leads to non physical solutions.  In the next subsections, we will see that this problem can be solved by considering more general metric functions. 

We conclude that in this case the corresponding interior line element can be expressed as
\begin{eqnarray} \nonumber
&& ds^2 = e^{2\nu} (1+qa)dt^2 - (1-qa) \bigg(\frac{dr^2}{1-\frac{2\tilde m}{r}} \\ 
&&+ r^2 d\theta^2 + r^2 \sin^2 \theta d\varphi^2\bigg) \ ,
\end{eqnarray}
with
\begin{equation}
a= \ln\left( 1- \frac{2m}{R } \right)\ .
\end{equation}
In the limiting case $q\rightarrow 0$, we turn back to the interior Schwarzschild solution. 
\subsection{\label{sec:rad}Solutions with spatial dependence}

We now assume that the functions $a$, $b$ and $c$ depend  on the radial coordinate. As before, the interior Schwarzschild solution 
(\ref{int_sch}) 
is taken as the zeroth approximation. An analysis of the field equations shows that the following cases need to be considered.

\textit{1)} Let $b(r)=0$ and $a(r)=c(r)$. The field equations allow only one solution, namely,  $a=const$. However, this is a trivial case that is equivalent to multiplying the density, pressure and some metric components by a constant quantity.

\textit{2)} Let $b(r)=0$ and $ a(r) \neq c(r)$. In this case, the boundary conditions (\ref{bouncona}), which imply that
$ a = c=a\left(R\right) $, are not consistent with the remaining field equations. No solutions are found in this case.

\textit{3)} Let $b(r) \neq 0$ and $ a(r) \neq c(r)$. From Eqs.~(\ref{eq4}) and (\ref{eq5}), we obtain that 
\begin{equation}
b(r)=-2a(r)+C_{ab} .
\end{equation}
where the $C_{ab}$ is a constant. This relationship simplifies the remaining equations. Nevertheless, we were unable to find analytical solutions. Therefore, 
we perform a numerical integration of the remaining equations for the functions $a(r)$, $c(r)$, and $p_1(r)$. 
We take as
a particular example the values specified in  (\ref{con_parmet}) for the mass $m$, the matching radius $r_{\Sigma}$ and the quadrupole parameter $q$. 
Then, from the boundary conditions (\ref{bouncona}), we obtain 

\begin{eqnarray}
&& a\left(r_\Sigma \right)=-1.529, \ c\left(r_\Sigma \right)=-2.715, b\left(r_\Sigma \right)= 3.0696\quad
\\ && a_{,r}\mid_{r=r_\Sigma}=25 \ ,C_{ab}=0.01.
  \label{boundvalues}
\end{eqnarray}

	{Moreover, to perform the numerical integration, it is necessary to specify the profile of the density function 
$\rho_1(r)$, which we take as
\begin{equation}
\rho_1(r) = \rho_1(0)-r - \frac{1}{2}r^2 - \frac{1}{6}r^3\ ,
\label{rho1r}
\end{equation}
 where the constant $\rho_1(0)$ must be chosen such that the total density $\rho = \rho_0+q\rho_1(r)$ vanishes at the surface and is finite at the center $r=0$. For the particular parameter values (\ref{con_parmet}), we obtain $\rho_1(0)=-8.718$ and the behavior of the total density is illustrated in Fig.~\ref{fig2}.}
\begin{figure}
\includegraphics[width=6.5cm,height=6cm,scale=0.2]{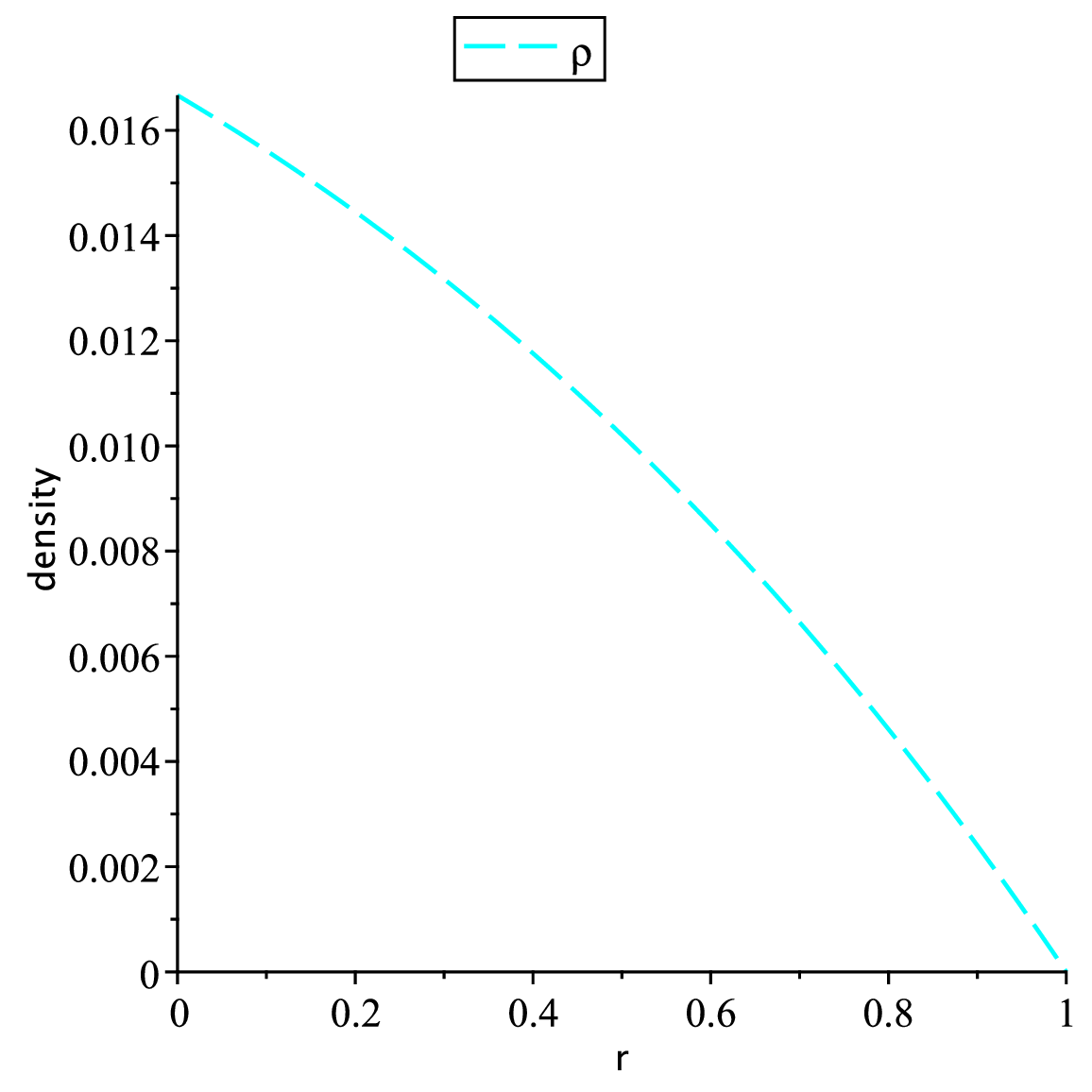}
\includegraphics[width=6.5cm,height=6cm,scale=0.2]{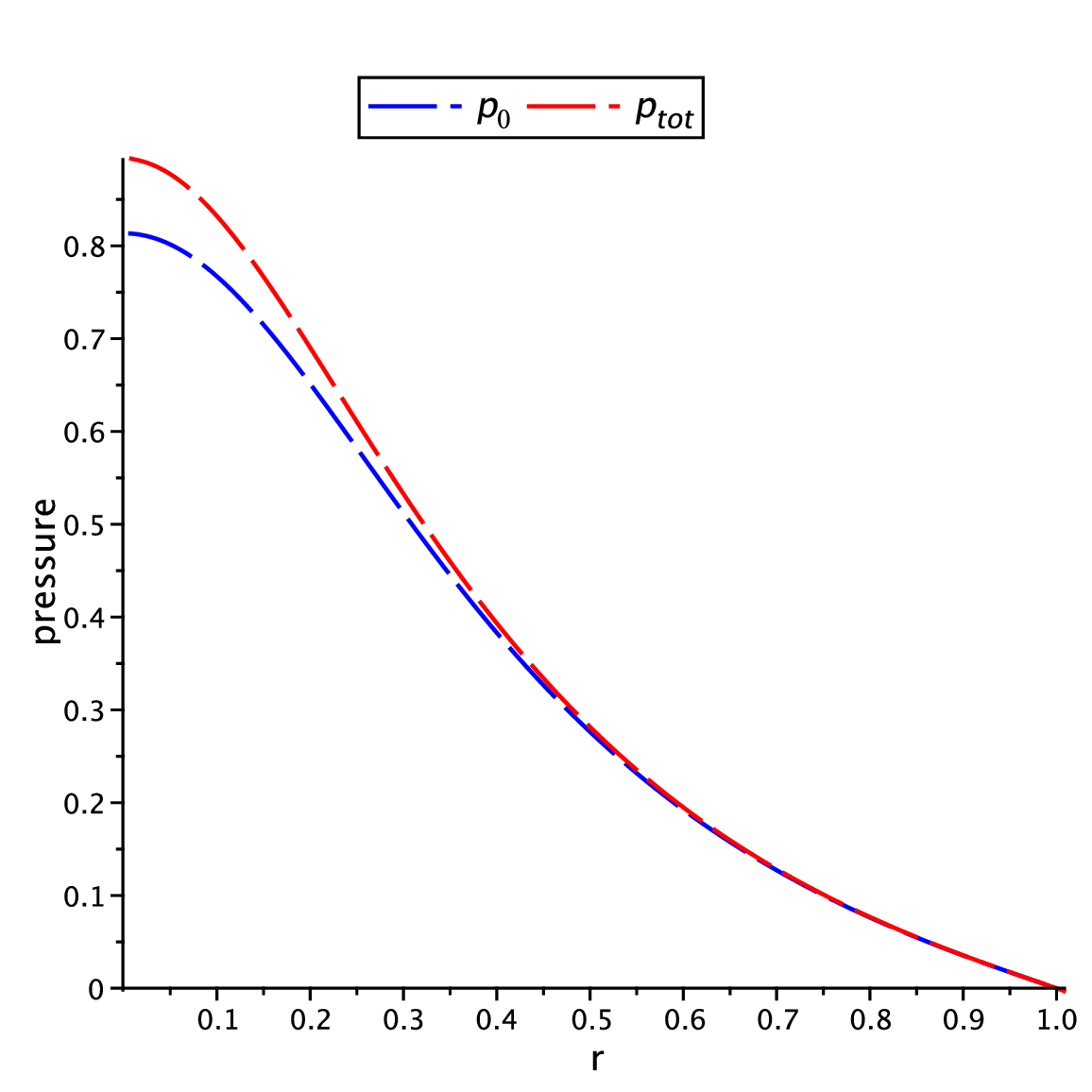}
\caption{Behavior of the density and pressure as functions of the radial coordinate.
\label{fig2}
}
\end{figure} 
With this density function, the numerical integration can be performed explicitly, 
leading to a solution for the pressure which is presented in Fig.~\ref{fig2}. 

Moreover, the result of the numerical integration of the metric functions 
$ a(r), b(r) $ and $c(r)$ is represented in Fig.~\ref{fig3}.
\begin{figure}
\centering
\includegraphics[width=8.3cm,height=6.5cm,scale=0.4]{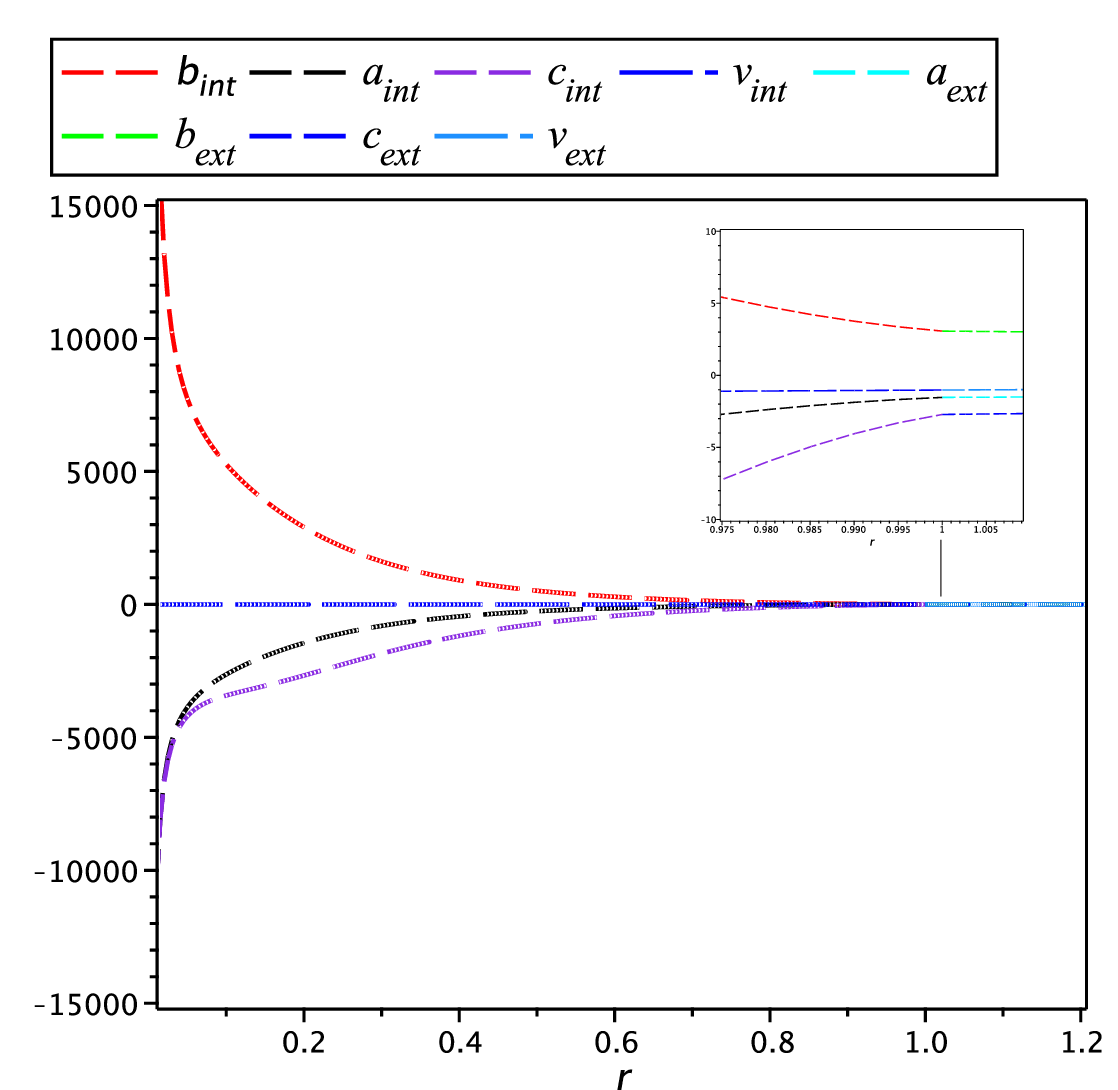}
\caption{Behavior and matching of the metric functions.
\label{fig3}
}
\end{figure}
We see that all the boundary conditions for the variables of the perfect fluid and the metric functions are satisfied and that all the quantities show a regular behavior.
Moreover, 
to further analyze the physical significance of this solution, we can determine the corresponding equation of state from the value of the total density and pressure (see Fig. \ref{fig2}). The result is shown in Fig. \ref{fig4}. We notice a realistic behavior as the pressure increases with the density. 
\begin{figure}
	\centering
	\includegraphics[width=8.5cm,height=6cm,scale=0.4]{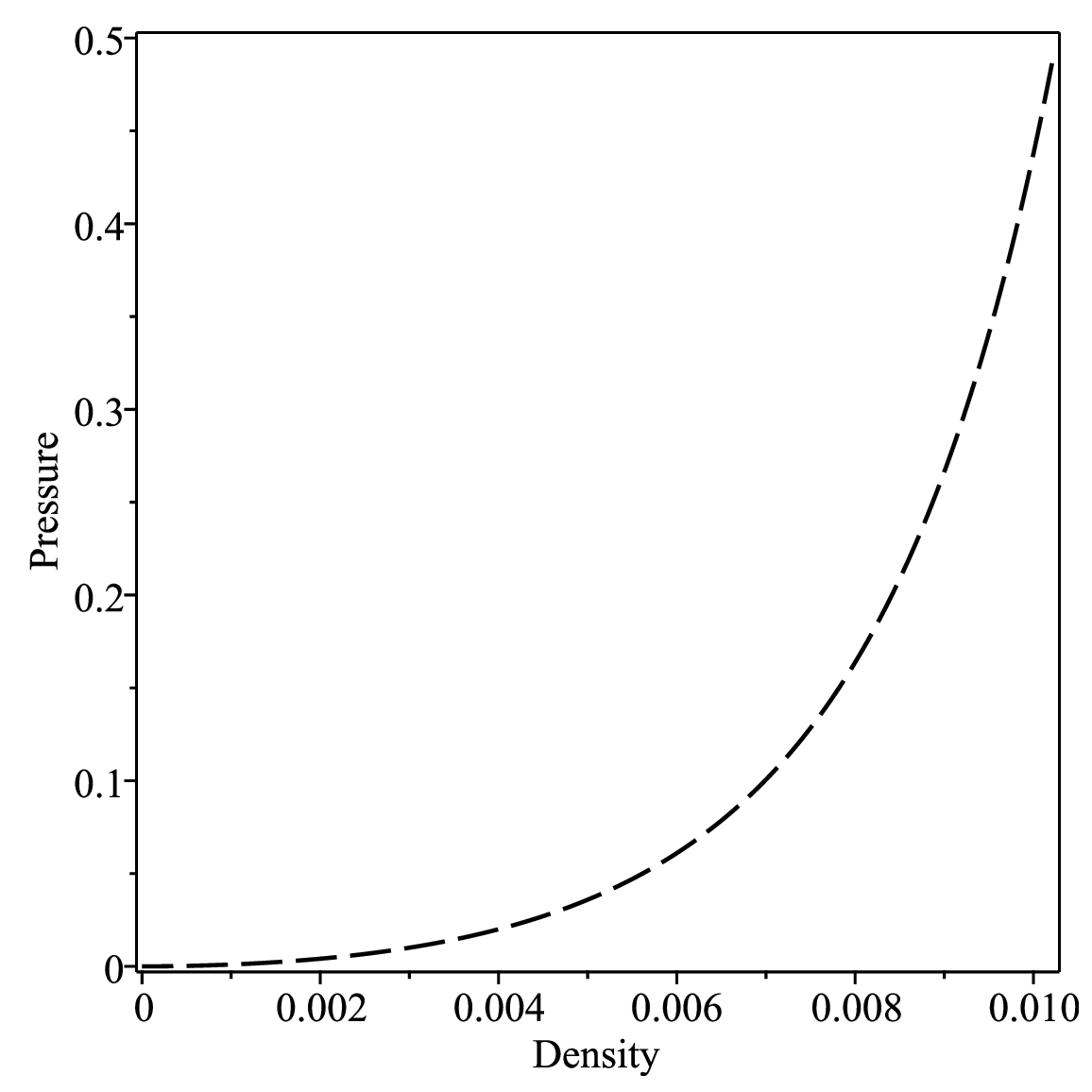}
	\caption{ Equation of state of the solution represented in Fig. \ref{fig2} ($P$ versus $\rho$ in geometrized units).
		\label{fig4}
	}
\end{figure}

We conclude that in this particular case it is possible to obtain physically meaningful solutions. For completeness, we mention that
the line element for this perfect fluid solution can be written as
\begin{eqnarray} \nonumber
ds^{2}&=& e^{2\nu} (1 + q a) dt^{2} - (1+qb+qc) \frac{dr^2}{1-\frac{2\tilde m}{r}}
\\ && - (1-qa)(d \theta^{2}+ r^{2} \sin^{2} \theta d \varphi^{2})
\end{eqnarray}
which, as expected, leads to the interior Schwarzschild spacetime for vanishing quadrupole parameter.

\subsection{Barotropic solutions}
\label{sec:bar}

A further method to obtain approximate interior solutions is to specify   {\it a priori} an EoS that relates density and pressure. 
 An interesting class of fluids are the barotropic fluids \cite{anile1990}, which   obey the EoS  
$    p=p \left( e \right)$,
{where $e$ is the total energy density.
One of the interesting cases of barotropic fluids is provided by matter at zero temperature. In this case, the barotropic EoS  is given as} 
$p=p \left( \rho \right)$ 
and 
$  e=e \left( \rho \right)$.
One of the simplest cases is represented by the barotropic relation 
\begin{equation}
p = w \rho \left(r \right) \ ,
\end{equation}
where $w$ is the barotropic constant factor. The corresponding field equations can be obtained by replacing the above equation in the equations presented in \ref{app_linfield}. To proceed with the numerical integration of the field equations in this case, we choose the free parameters as
\begin{eqnarray}
\label{con_parmet2}
&&\rho_0= 0.7, \ r_{\Sigma}=1,\ m=0.23626,\  q=\frac{1}{100},C_{ab}=0.01,
 \nonumber
\\ 
&& \nu(r_{\Sigma} ) =-0.3198, \ a(r_{\Sigma} )=-0.6186, \  c(r_{\Sigma}) =-0.8101.
\end{eqnarray}                                                                                                          
Moreover, we assume the boundary conditions 
(\ref{bouncona}) for the metric functions and the vanishing of the pressure (\ref{bounconp}) on the matching surface. The boundary conditions are then used as initial values for the numerical integration. The behavior of the resulting total density is represented in Fig.\ref{fig5} for $w=0.3$.
\begin{figure}
	\includegraphics[width=8.5cm,height=7cm,scale=0.41]{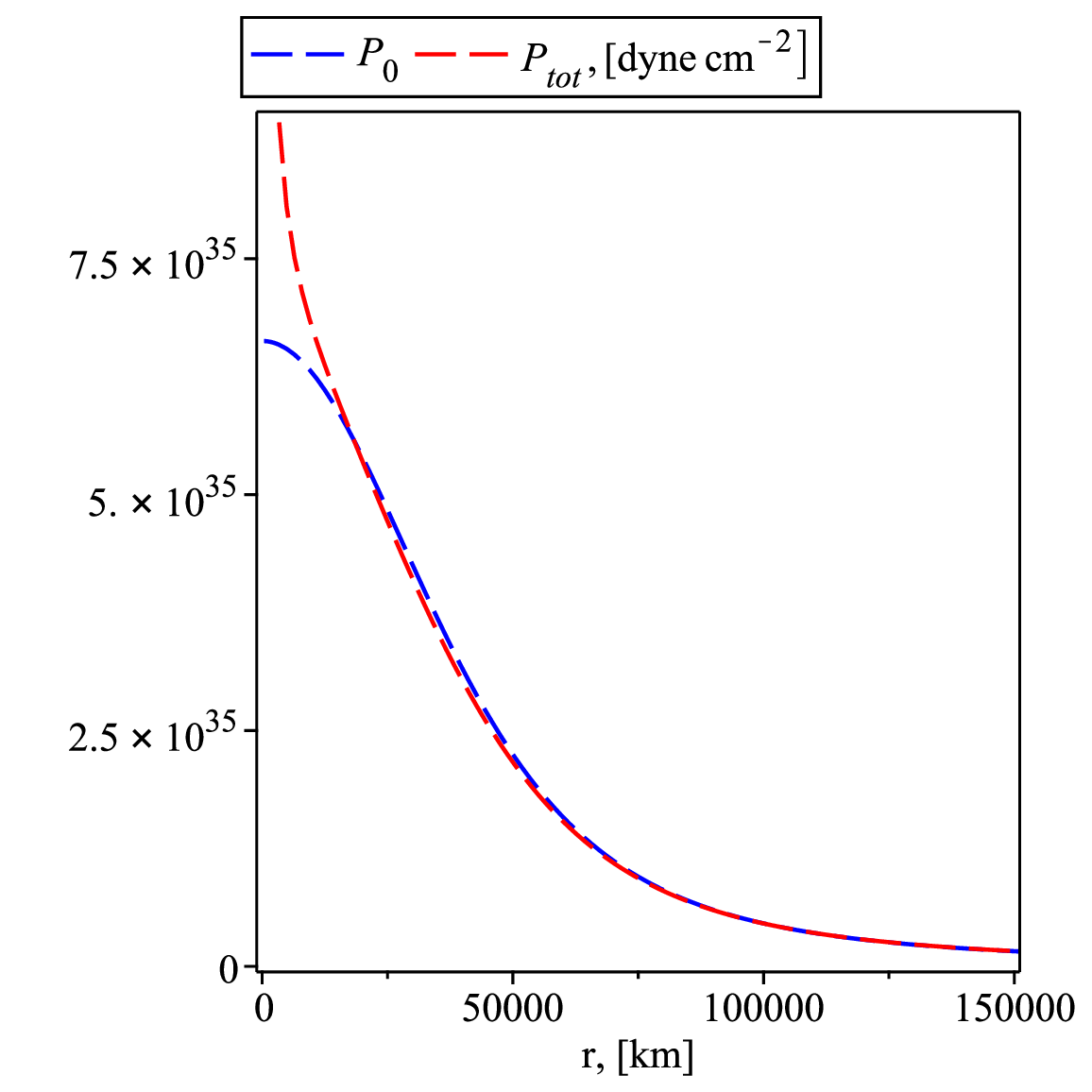}
	\caption{Behavior of the total density in cgs units. The central density is $\rho_{0}=10^{6}$ $g/cm^{3}$
		\label{fig5} 
	}
\end{figure}
To test our approach, we calculate numerically the value of the total pressure and obtain a behavior similar to that of the total density, which confirms the validity of the barotropic EoS as shown in Fig. \ref{fig6}. 
\begin{figure}
	\includegraphics[width=8.5cm,height=6cm,scale=0.4]{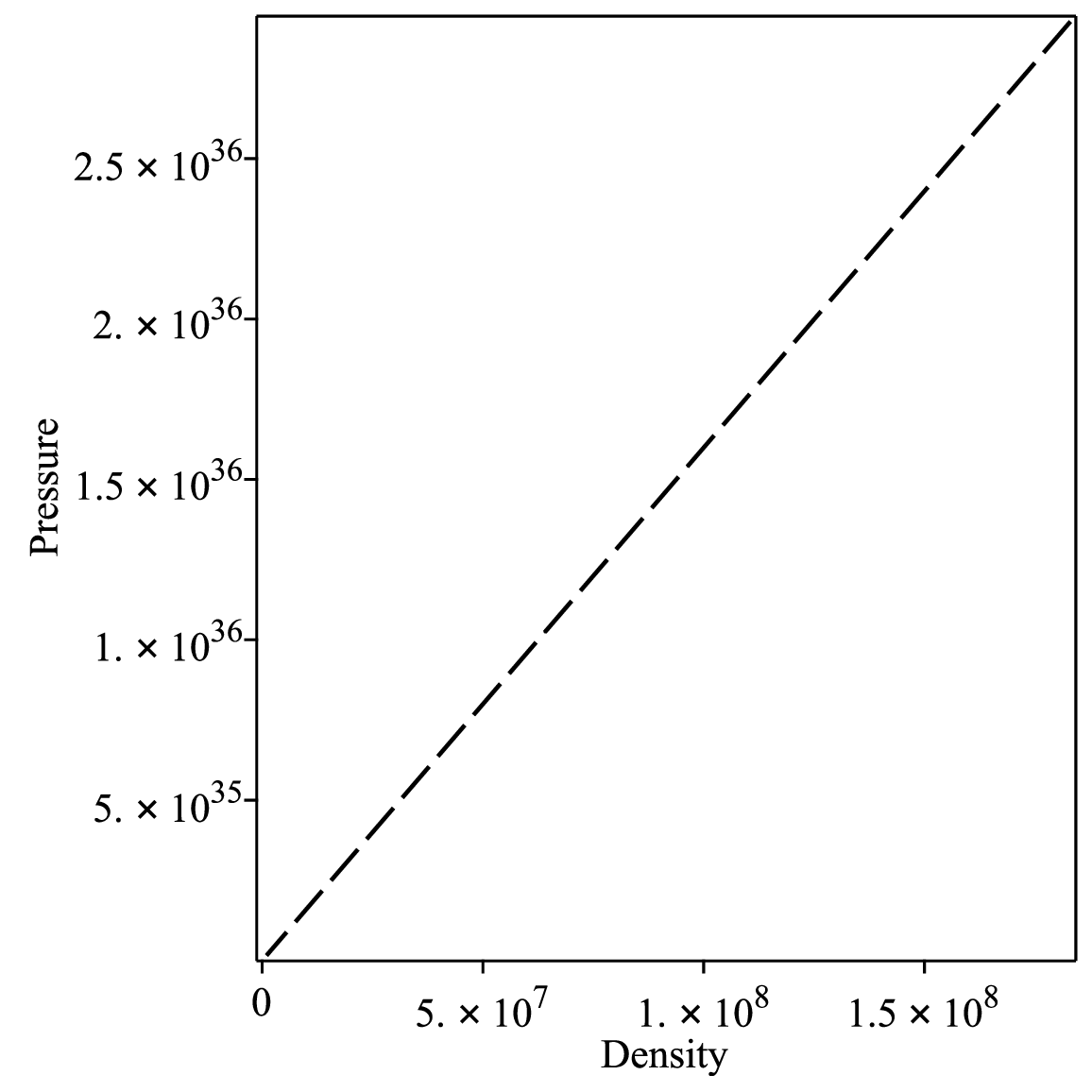}
	\caption{Barotropic behavior of  pressure versus density $P(\rho)$ in cgs units with $ w= 0.3 $.
		\label{fig6} 
	}
\end{figure}
Furthermore, the numerical integration of the field equations indicates that the metric functions are regular inside the source and can be matched with the metric functions of the exterior $q-$metric. This is shown explicitly in Fig. \ref{fig7}.
\begin{figure}
	\includegraphics[width=8cm,height=7cm,scale=0.35]{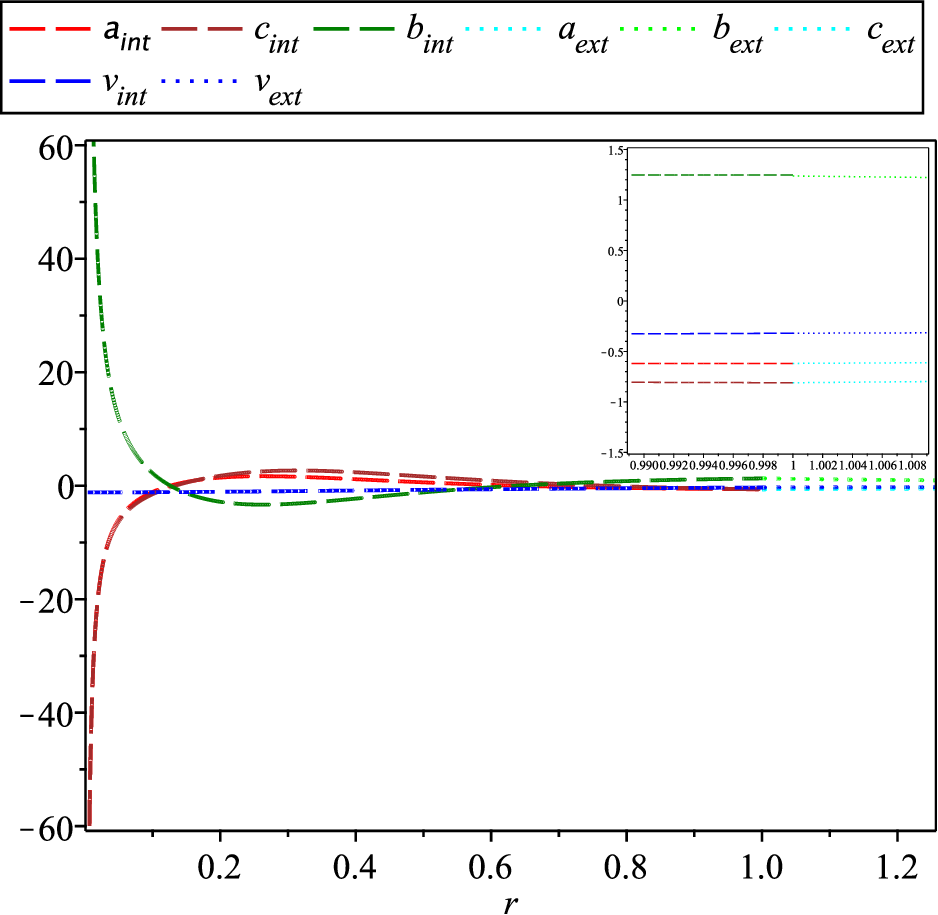}
	\caption{  Metric functions inside and outside the matching surface in geometrized units.
		\label{fig7}
	}
\end{figure}

We conclude that it is possible to obtain approximate interior solutions with a barotropic EoS, which can be consistently matched with the exterior metric functions and are characterized by  well-behaved density and pressure functions. 

\subsection{Polytropic solutions}
\label{sec:poly}

A polytrope refers in the case considered here to a perfect fluid in which the pressure depends upon the density in the form 
\begin{equation}
p= k \rho^\gamma\ ,
\end{equation}
where $k$ is a proportionality constant and $\gamma$ is the polytropic index. The corresponding field equations can be integrated numerically by assuming the values (\ref{con_parmet2}) for the free parameters.

Using the boundary conditions (\ref{bouncona}) and assuming the vanishing of the pressure on the matching surface, we obtain for the pressure the behavior depicted in Fig.\ref{fig11}, whereas the metric functions that satisfy the field equations and the matching conditions can be represented as in Fig.\ref{fig10}.
	
	\begin{figure}
	\centering
	\includegraphics[width=8.5cm,height=6cm,scale=2]{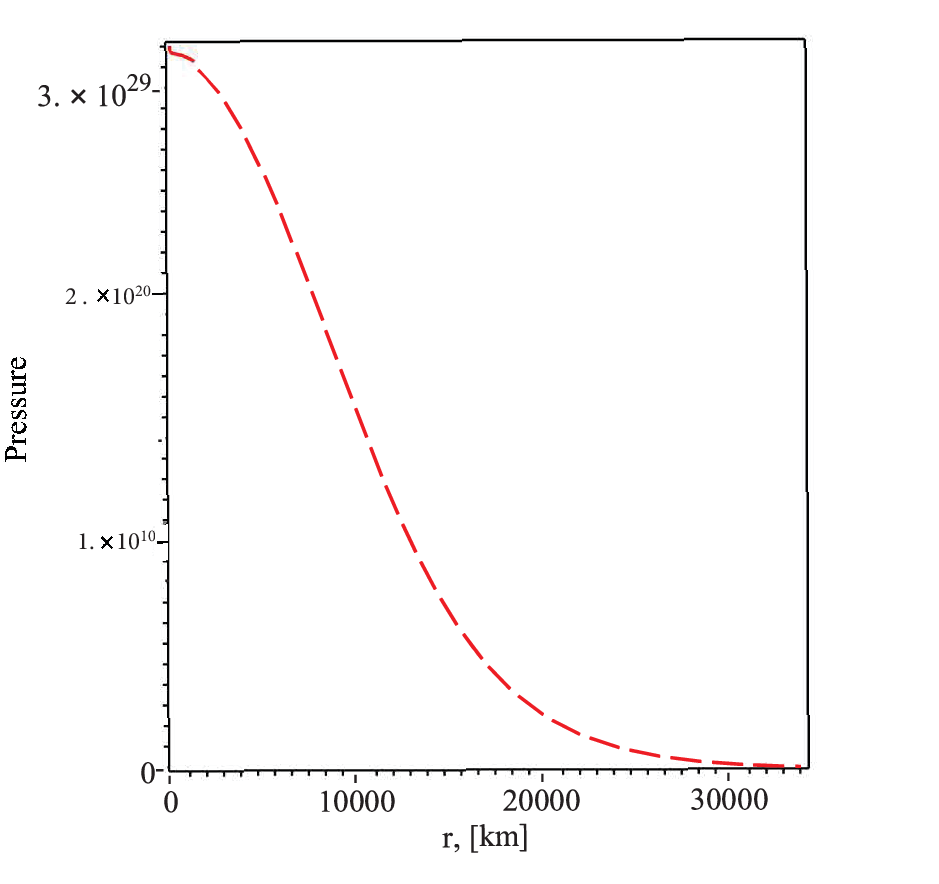}
	\caption{Behavior of the total pressure in cgs units for a polytropic perfect fluid with  $k=0.43$ and  $\gamma= 3/2$.}
		\label{fig11}
	\end{figure}

\begin{figure}
	\centering
	\includegraphics[width=8cm,height=6cm,scale=0.6]{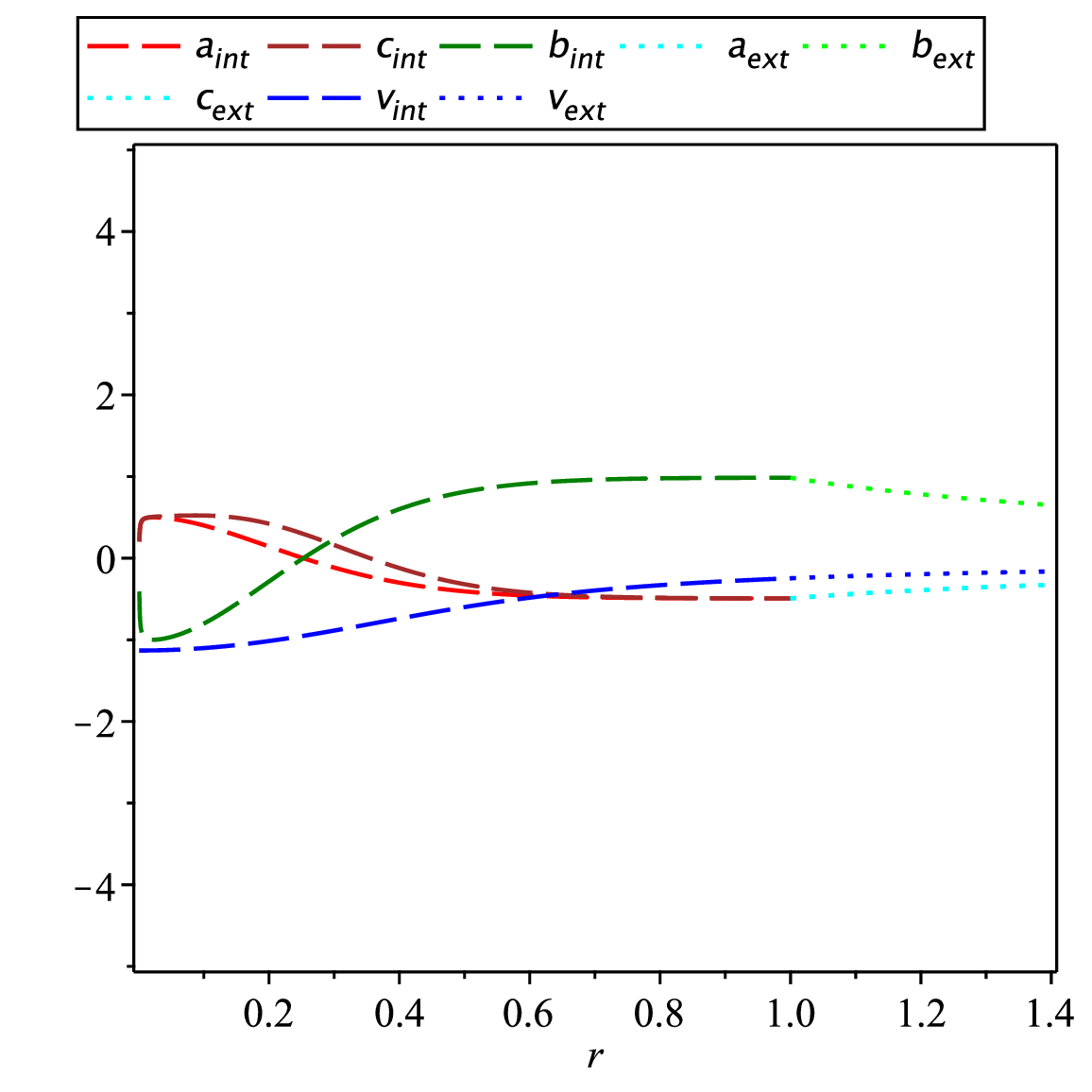}
	\caption{Behavior of the metric functions for a  polytropic perfect fluid with $k=0.43$  and  $\gamma= 3/2$. }
		\label{fig10}
	\end{figure}

Thus, we see that in the case of a polytropic fluid our approach allows us to obtain approximate interior solutions that can be matched with the exterior $q-$metric. The particular values of the proportionality constant $k=0.43$ and the polytropic index $\gamma=3/2$ have been chosen such that they correspond effectively to a realistic physical system, as will be shown in the next section.

\section{On the physical significance of the solutions}\label{sec:phys}

To derive interior solutions with quadrupole moment that can be matched with the exterior approximate $q$-metric, first we have made a series of assumptions that reduce the mathematical complexity of the underlying field equations. As a result, we obtained numerical solutions that show the mathematical consistency of the proposed method. However, the physical significance of the solutions and their applicability to describe realistic compact objects seem to be drastically reduced by the assumption of the before mentioned mathematical conditions.  Indeed, the first solution we obtained in Sec. \ref{sec:fcons} is characterized by a constant value of the energy density and constant values of some interior metric functions which, as mentioned before, lead to discontinuities in the derivatives of the metric functions and, consequently, to nonphysical solutions.

In Sec. \ref{sec:rad}, we applied a different procedure in which the profile of the density is given {\it a priori} as a function of the spatial coordinate. In this case, we obtain numerical solutions for the pressure and  metric functions with a more realistic physical behavior, as shown in Figs.~\ref{fig2} and \ref{fig3}. Indeed, the density and pressure are maximal at the center of the source, then their value diminish continuously until they vanish at the surface of the source. 

Then, in Sections \ref{sec:bar} and \ref{sec:poly}, we apply a different method that consists in  specifying  {\it a priori} an equation of state, which allows us to integrate the corresponding field equations. We assume 
barotropic and a polytropic equations of state and find numerical solutions for the pressure and the metric functions that are well behaved from a physical point of view.   

We now investigate whether it is possible to apply our results to describe the gravitational field of realistic compact objects,  in which several properties of the internal structure of the gravitational sources are taken into account. Indeed, the most recent surveys with observational data show that white dwarfs
\cite{2019MNRAS.486.2169K,2019MNRAS.482.4570G,2020ApJ...898...84K}
and neutron stars
\cite{2016EPJA...52...63M,2019ApJ...874...64Z,2019ApJ...887L..27G,2019ApJ...887L..25B,2020PhRvD.101l3007L}
are realistic compact objects, where relativistic effects are expected to play a non-negligible role. 
Accordingly, we will compare the physical properties of our solutions, as expressed at the level of the EoS, with those of realistic compact objects.
Consider, for instance, a white dwarf whose interior is described by the Chandrasekhar EoS in  parametric form (geometrized units) 
\cite{zn71,chandrasekhar31,1971tges.book.....Z,shapirobook}

\begin{eqnarray}
\rho_{Ch}& =&\frac{32}{3} \left(\frac{m_e}{m_n} \right)^{3} K_{n} \left( \frac{ \overline{A}}{Z} \right)  y\left(x\right)^3\ , \label{EoS1_Ch_WD}
\end{eqnarray}

\begin{eqnarray}
p_{Ch}&=& \frac{4}{3} \left(\frac{m_e}{m_n} \right)^{4} K_{n} \bigg[ 
y \left(x \right) \left(2 y \left(x \right)^{2}-3 \right) \sqrt{1+y\left(x \right)^{2}} \nonumber \\ && + 3 \ln \left(y \left(x \right)+ \sqrt{1+y\left(x \right)^{2}} \right)\ ,
\bigg]
\label{EoS2_Ch_WD}
\end{eqnarray}
with
\begin{equation}
K_n= \frac{m^{4}_{n} }{32 \pi^{2}  },
\end{equation}
where
$ \overline{A}$ and $Z$ are the
average atomic weight and atomic number of the corresponding nuclei; $ y \left(x \right)= p_{e} \left(x \right) / m_{e}   $, with $p_{e} \left( x \right)$  , $m_{e}$ and $ m_{n}$ are the Fermi momentum, the mass of the electron and the mass of the nucleon, respectively. Here, we consider the particular case for the average molecular weight $\overline{A}/Z = 2$. 

The Chandrasekhar EoS is considered to be the simplest,  but also the most relevant and basic EoS for the description of white dwarf matter. It should be noted that there are plenty of more sophisticated EoS, which are used to describe the interior of white dwarfs and outer crusts of neutron stars
\cite{1961ApJ...134..669S,2011PhRvD..84h4007R,2020EPJP..135..290C,2019MNRAS.490.5839B,haenselbook}. These equations take into account the electron-electron, electron-ion and ion-ion Coulomb interactions, nuclear composition, Thomas-Fermi corrections,  finite temperature effects,  phase transitions, magnetic fields, etc.
\cite{1990RPPh...53..837K,sheyse,2020arXiv200400846B,b2016b,boshmg18,2020EPJP..135..290C,Faussurier2017,2017JSMTE..11.3101F}. However, for our purposes we will restrict ourselves to the Chandrasekhar EoS for simplicity and clarity.

In the case of neutron stars, the internal structure can be described by the EoS of a pure degenerate neutron gas, which in parametric form can be written as \cite{zn71,shapirobook,haenselbook} 
\begin{eqnarray} \nonumber
\rho_{NS}&=&\frac{\epsilon_{0}}{8} \bigg[ \left(2 y \left(x \right)^{3}+y \left(x \right) \right) \sqrt{1+y\left(x \right)^{2}}  \\ && - \ln \left(y \left(x \right)+ \sqrt{1+y\left(x \right)^{2}} \right)
\bigg], \\ \nonumber
p_{NS}& =& \frac{\epsilon_{0}}{24} \bigg[ 
\left(2 y \left(x \right)^{3}-3y \left(x \right) \right) \sqrt{1+y\left(x \right)^{2}} \\ &&+ 3 \ln \left(y \left(x \right)+ \sqrt{1+y\left(x \right)^{2}} \right)
\bigg],
\end{eqnarray}  
where $\epsilon_{0}= m_{n}^{4} c^{5}/ \pi^2 $ is the energy density.

It should be noted that the pure degenerate neutron gas represents the simplest EoS for neutron stars, though there is a plenty of other more complicated EoSs in the literature, accounting for the nucleon-nucleon interactions, contribution of various particles-carriers of  interaction, etc. \cite{haenselbook,2014NuPhA.921...33B}. The most recent and realistic EoSs are tested via measuring X-ray emissions, tidal deformation and gravitational wave events   \cite{2020AAS...23523703R,2020ApJ...893L..21R,2020arXiv200400846B,2020arXiv200704427B,2020arXiv200603168C,2020MNRAS.495.5027P,2020arXiv200514164G,2020arXiv200407232D}. 

To prove that the interior metrics investigated in this work can be used to describe the internal gravitational field of white dwarfs and neutron stars, we should integrate the above EoSs together with the field equations given in \ref{app_linfield}. A preliminary computation shows that this is possible, but requires a detailed analysis of the parameters and constants that enter the EoSs. This we expect to investigate in future works.  For the purposes of the present work, we propose to use a different method, which consists in finding an effective EoS from the above parametric EoS. To this end, we compute the parametric pressure and density and plot the result as an effective EoS of pressure vs. density. The result is shown in Fig. \ref{fig8} for white dwarfs and in Fig. \ref{fig9} for neutron stars, respectively.
\begin{figure}
	\centering
	\includegraphics[width=8.5cm,height=6cm,scale=0.6]{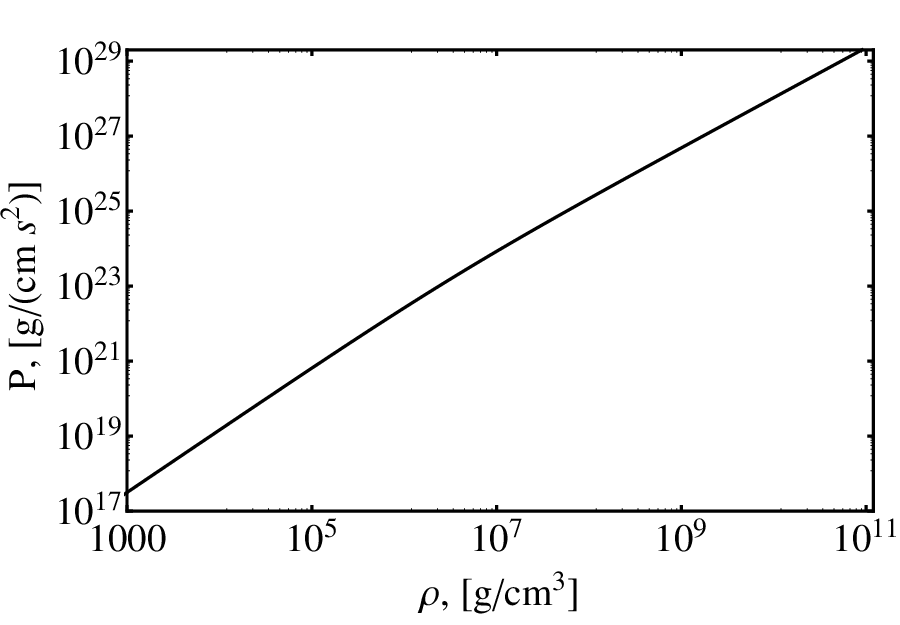}
	\caption{ Effective EoS (log scale) for white dwarfs obtained from the parametric Chandrasekhar EoS.  }
		\label{fig8}
	
\end{figure}
\begin{figure}
	\centering
	\includegraphics[width=8.5cm,height=6cm,scale=0.6]{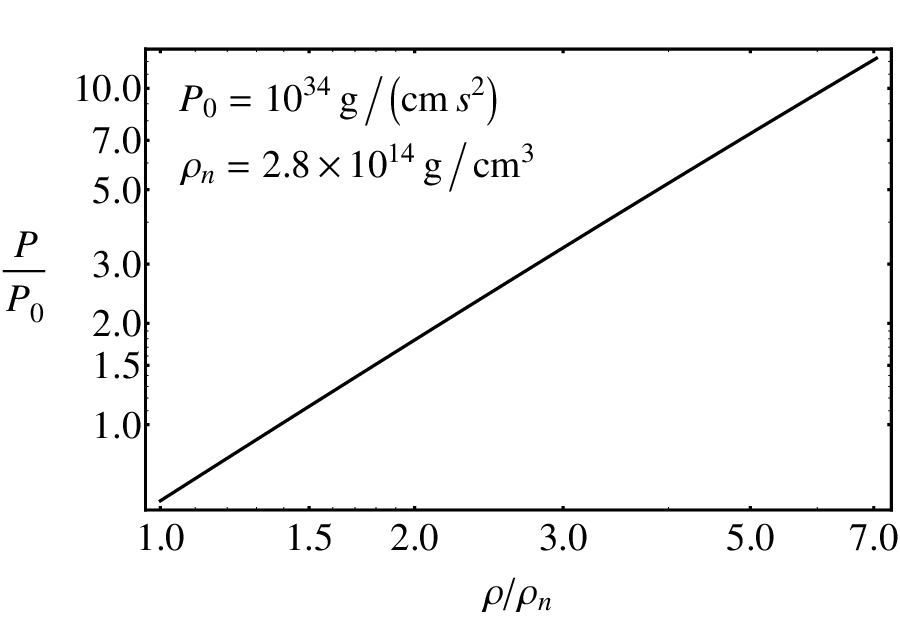}
	\caption{Effective EoS (log scale) for neutron stars described by a pure degenerate neutron gas. Here $\rho_n$ is the average nuclear density.  }
		\label{fig9}
\end{figure}
It is then easy to show that in both cases the effective EoS can be approximated by a polytrope, $p=k \rho^\gamma$, with $k=5.5 10^{-6}$ and  $\gamma =3/2$ for white dwarfs and $k=7.13 10^{19}$ and $\gamma= 1.365$ for neutron stars, respectively. On the other hand,  in Sec. \ref{sec:poly}, we have shown that it is possible to obtain interior solutions with a polytropic EoS. We thus conclude that the method presented in this work can be used to describe the interior gravitational field of white dwarfs and neutron stars with an effective polytropic EoS.

However, notice that to integrate the field equations in our approach we must fix {\it a priori} the values of the mass and the matching radius so that we cannot establish any limits on these quantities, which are very important for the understanding of the physical properties of astrophysical compact objects like white dwarfs and neutron stars. Our approach does not allow us to investigate this problem. To this end, it is necessary to take into account the equilibrium conditions for the interior solution. In the case of spherically symmetric interior solutions, this is done by considering the Tolman-Oppenheimer-Volkoff (TOV) equation. Consequently, to perform a more physical analysis of the inner structure of astrophysical objects with quadrupole moment, it is necessary to generalize the TOV procedure to include axially symmetric sources. We expect to investigate this problem in future works.

\section{\label{sec:con}Conclusions and remarks}

In this work, we have investigated interior solutions of Einstein's equations in the case of static and axially symmetric perfect fluid spacetimes. We impose the physical condition that the interior spacetime can be matched smoothly with the exterior $q-$metric characterized by two parameters, which determine the mass and quadrupole moment of the source. 

 We assume in this work that the interior counterpart of the exterior $q-$metric is described by an isotropic perfect fluid, for the sake of simplicity. Indeed, as has been shown in \cite{hdio13}, the most general interior solution, which is compatible with an exterior static and axisymmetric gravitational source, accepts up to four stresses. Thus, we are assuming here that three of them are negligible small so that we end up with a perfect fluid with only one isotropic stress. Moreover, the isotropic perfect fluid is a very idealized model since it has been shown that static and isotropic perfect fluid sources must be spherical, at least in the case of an incompressible equation of state \cite{masood09,hdio13}. This is related to the fact that even small pressure anisotropies can generate breaks in the fluid distribution, drastically moving away from the idealized isotropic configuration \cite{herrera92}. Furthermore,  analyzing the stability of the isotropy condition,  it has been established recently in \cite{herrera2020} that the realistic physical processes that  occur during the stellar evolution inevitably lead to the appearance of pressure anisotropies, which cannot disappear during the dynamic evolution of the mass distribution. Consequently, several results indicate that the final equilibrium configuration of a realistic stellar evolution is characterized by the presence of pressure anisotropies. Nevertheless, in this work, we assume the isotropy condition to simplify the mathematical complexity of the resulting field equations. 
 Our results show that it is possible to find approximate perfect fluid solutions with quadrupole, which are consistent from the mathematical point of view, in the sense that Einstein's equations are satisfied for certain equations of state.
 
 In general, it is difficult to find interior solutions for a given exterior solution by using Einstein's equations. 
Due to the complexity of the corresponding set of differential equations, 
we reduce the problem to the particular case in which the quadrupole parameter can be considered as a small quantity, which is then used to linearize the exterior $q-$metric. As for the interior metric, we use the particular linearized Ansatz (\ref{apin1}), leading to a simplified set of differential equations in which, as a consequence of the conservation law (\ref{eqp1}) and (\ref{rhotheta}), the parameters of the perfect fluid do not depend on the angular coordinate $\theta$. 
We then analyze the corresponding linearized field equations and derive several classes of new vacuum and perfect fluid solutions, which depend on the spatial coordinate $r$, only. 

We search for interior solutions that  can be matched with the approximate exterior $q-$metric, satisfy the energy conditions for the density and the pressure and are free of singularities in the entire spacetime.  Our results show that it is possible to find such solutions, implying that the approximate approach is compatible with the imposed physical conditions. 
Indeed, we first found a particular analytical solution for which pressure and density are well-behaved functions, but the derivatives of the metric functions present discontinuities that are considered as nonphysical. However, when we impose a particular functional dependence for the density, a barotropic or a polytropic  EoS, we obtain numerical solutions that satisfy the matching conditions for the metric functions and the energy conditions for the fluid. In addition, we analyzed the behavior of the pressure and density as given at the level of the EoS.  To verify if these particular solutions can be applied to describe the exterior and interior gravitational field of realistic compact objects, we analyzed the Chandrasekhar EoS for white dwarfs and the pure degenerate neutron gas EoS for neutron stars. We found that the inner properties of these compact objects can be represented effectively by a polytropic EoS, which is essentially the behavior observed in the numerical solutions found in Sec. \ref{sec:poly}.  We interpret this result as an indication that our approximate solutions with arbitrary mass and small quadrupole moment can be used to study the exterior and interior gravitational field of relativistic compact objects. Nevertheless, we also noticed that the model developed in this work does not impose any limits on the value of the mass of compact objects, a result which is well known from observations and more sophisticated theoretical studies. This is a disadvantage of our model that can be improved by considering the equilibrium conditions for the interior solutions. To this end, it would be necessary to consider a generalization of the TOV procedure to include axisymmetric gravitational fields. We will consider this problem in future works. 

In the literature, there are several methods that deal with approximate solutions of Einstein's equations \cite{hartle1967,ht1968,Hernandez1967,1999A&A...352..211K,2017IJMPS..4560029Z} 
Usually, they intend to find numerical solutions, starting from approximate  metrics, field equations, numerical EoS and sometimes even approximate coordinates. Sometimes those methods do not care about the matching conditions at the level of the derivatives of the metric.  With the same initial conditions and physical assumptions, the method we use here should be able to reproduce the results obtained by other methods. However, the intention of our method is to go beyond approximate numerical solutions. For instance, in the case of the quadrupole, other methods do not care about the form of the corresponding exact metric. In our case, this is very important because we are using the simplest known exact generalization of the Schwarzschild solution with quadrupole \cite{2018RSOS....570826F}, for which we expect to find a physically reasonable interior counterpart. The approximate solutions found in this work show that our method is self-consistent and can deliver physically meaningful solutions. This first approximate approach is necessary as a first proof of the capability of the method.

The obtained solutions are mathematically consistent, but are physically restricted because they depend on only one spatial coordinate. A more realistic case would necessarily imply to consider solutions with an additional angular dependence, which corresponds to taking into account the rotation of the gravitational source. To this end, it will be binding to investigate generalizations of the particular Ansatz (\ref{apin1}). This is a task of a future work. 

The results obtained here can be considered as a first step towards the determination of an exact solution of Einstein's equations which describes correctly the gravitational field of a rotating deformed source. The important feature of our approach is that we consider explicitly the influence of the quadrupole on the structure of spacetime and the corresponding field equations. In the present work, we only considered the simple and idealized case
of a static mass distribution with a small quadrupole and obtained compatible and physically 
reasonable results. To study more realistic configurations, it is necessary to take into account the rotation and the exact quadrupole of the mass. We expect to investigate these problems in future works.

\section*{Acknowledgments}

This work was partially supported  by Ministry of Education and Science (MES) of the Republic of Kazakhstan (RK), Grant No. BR10965191, and by UNAM-DGAPA-PAPIIT, Grant No. 114520, 
Conacyt-Mexico, Grant No. A1-S-31269. 
K.B. acknowledges the MES of the RK for support, Grant IRN: AP08052311. S.T. acknowledges the support through the postdoctoral fellowship program of Al-Farabi Kazakh National University. 


\appendix

\section{The HPHM metric}
\label{app:hphm}

Hern\'andez-Pastora, Herrera and Marti (HPHM) presented recently in \cite{Hernandez_Pastora_2016}  a general procedure to find static and axially symmetric interior solutions to the Einstein equations by focusing on the matching of the interior spacetime to a given exterior solution. To this end, a particular line element was chosen that can be expressed as 
\begin{equation}
ds^2=-e^{2 \hat{a} } Z \left(r\right)^{2}dt^2= \frac{e^{ 2 \hat{g}-2 \hat{a}}} {A\left( r\right)}dr^2+e^{ 2 \hat{g}-2 \hat{a}} r^2 d \theta^2 +e^{ 2 \hat{a}}r^2 \sin^2 \theta  d \varphi^2\ ,
\label{her_line}
\end{equation}
where $\hat a$ and $\hat g$ are functions of $r$ and $\theta$, in general. 

To find the relationship between the interior solution found in \cite{Hernandez_Pastora_2016} for the exterior $q-$metric and the interior solutions presented in the present work, we compare (\ref{her_line}) with the line element proposed here in Eq.(\ref{apin1}), i.e.,
\begin{eqnarray}
&& ds^2 =  e^{2 \nu} (1+qa) dt^2 - (1+qc+qb)\frac{dr^2}{1-\frac{2\tilde m}{r}}  \nonumber \\ &&- (1+qa+qb)r^2d\theta^2
-(1-qa)r^2\sin^2\theta d\varphi^2\ .
\end{eqnarray}
Since the set of coordinates is the same in both line elements, we can perform a direct comparison term by term of the metric functions and obtain the set of equations
\begin{equation}
e^{2\hat a} Z^2 = e^{2\nu} (1+qa) \ ,
\label{com1}
\end{equation}
\begin{equation}
e^{2\hat g - 2\hat a } A^{-1} = (1+qc+qb)\left(1- \frac{2\tilde m}{r} \right) ^{-1}\ ,
\label{com2}
\end{equation}
\begin{equation}
e^{2\hat g - 2\hat a } =  1+qa+qb \ ,
\label{com3}
\end{equation}
\begin{equation}
e^{- 2\hat a }  = 1-q a\ .
\label{com4}
\end{equation}
Obviously, the compatibility  between Eqs.(\ref{com1}) and (\ref{com4}) is guaranteed. On the other hand, since the functions $A$ and $\tilde m$ depend on $r$ only and both of them can be absorbed by an appropriate reparametrization of the  coordinate $r$, without loss of generality we can set
\begin{equation}
A = 1 - \frac{2 \tilde m}{r} \ .
\end{equation}
Then, from Eqs.(\ref{com2}) and (\ref{com3}), we obtain
\begin{equation}
e^{2\hat g - 2\hat a }  = 1+qc+qb = 1 + qa + q b\ ,    
\end{equation}
which implies that 
\begin{equation}
    a = c\ .
\end{equation}
All the physically relevant interior solutions that we find in Sec. \ref{sec:sol} are characterized by the condition $a\neq c$.  It follows that the interior metrics  derived in the present work cannot be obtained as particular solutions of the HPHM solution.


\section{Linearized field equations}
\label{app_linfield}

In general, up to the first order in $q$, the field equations which follow from the line element 
\begin{eqnarray}
&& ds^2 =  e^{2 \nu} (1+qa) dt^2 - (1+qc+qb)\frac{dr^2}{1-\frac{2\tilde m}{r}}  \nonumber \\ &&- (1+qa+qb)r^2d\theta^2
-(1-qa)r^2\sin^2\theta d\varphi^2\ ,
\end{eqnarray}
where all the metric functions depend on $r$ only, 
can be written as
\begin{equation} 
\tilde m_{,r} = 4 \pi \rho_0 r^2 ,
\end{equation}
\begin{equation}
\nu_{,r}=\frac{4 \pi p_{0} r^3 + \tilde{m}}{r \left(r-2 \tilde{m} \right)} ,
\end{equation},
\begin{equation}
p_{0,r}=-\frac{\left( 4 p_{0} \pi r^3+\tilde{m} \right) \left(p_{0}+ \rho_{0} \right)}{r\left(r-2 \tilde{m}\right)},
\end{equation}
\begin{eqnarray} \nonumber
&& 2r\left(r-2\tilde{m} \right)a_{,rr}
+ \left[ \left(3p_{0}-2 \rho_{0} \right)4 \pi r^3 +3 r -\tilde{m} \right]a_{,r} \\ && + \left(r-3\tilde{m} -4 \pi p_{0}r^3 \right) c_{,r}  
- 16 \pi r^2 \big[ \left(b+c \right)\left( \rho_{0}+p_{0}\right) \nonumber \\ && +\rho_{1}+p_{1}  \big] - 2 \left(a-c \right)=0,
\label{eq3}
\end{eqnarray}

\begin{eqnarray}
&&r \left( r-2 \tilde{m}\right) b_{,rr}  + \left(r-\tilde{m} - 4 \pi \rho_0 r^3 \right) b_{,r} \nonumber \\ && -2 \left(r-2\tilde{m} \right) c_{,r}+ 16 \pi r^2\big[ \left(c+b \right) \rho_{0} +\rho_{1} \big] \nonumber \\ && + 2 \left(a-c \right)=0,
\label{eq2}
\end{eqnarray}

\begin{eqnarray} \nonumber
&&\left( 4\pi p_{0} r^3 +r -\tilde{m} \right)
\left( a_{,r}- c_{,r}\right) 
-32 \pi r^2 
\big[ \left(c+b \right) p_{0} +p_{1} 
 \big] \\ && + 2\left( a-c \right)=0,
\label{eq1}
\end{eqnarray}

\begin{eqnarray} \nonumber
&&2 \left( 4\pi p_{0} r^3 + \tilde{m}\right) a_{,r}
+ \pi r^2 \big[\left( c+b\right)p_{0} \nonumber \\ && +p_{1}\big] -2 \left(a-c \right)  = 0,
\label{eq4}
\end{eqnarray}

\begin{eqnarray}  \nonumber
&&\big[ 4 \left( 4\pi p_{0} r^3 +r -\tilde{m} \right)^2 \sin^2 \theta +r \left(r-2 \tilde{m} \right) \cos^2 \theta  \big] b_{,r} \nonumber \\
&& + 2 \big[  \left( 4\pi p_{0} r^3 -\tilde{m} \right) \sin^2 \theta + r  \big]  \left(r-2 \tilde{m} \right) a_{,r} \nonumber \\
&& -2 \sin^2 \theta \left( 4\pi p_{0} r^3 +r -\tilde{m} \right) \big\{ 8 \pi r^2 \big[ \left( c+b \right)p_{0}+p_{1} \big] \nonumber\\
&& - a+c \big\}
=0.
\label{eq5}
\end{eqnarray}

The first three equations constitute a separated set of differential equations that can be integrated for any given value background density $\rho_0$. From this set of equations we can derive solutions for the metric functions $\nu$ and $\tilde{m}$ and for the background pressure $p_0$. These solutions can then be used in the second set of equations (\ref{eq3})-(\ref{eq5}) to find  solutions for the metric functions $a$, $b$ and $c$ and for the pressure function $p_1$. 

\bibliographystyle{ws-ijmpd}
\bibliography{app.bib}
\end{document}